\documentclass[aps,prd,preprint,nofootinbib,superscriptaddress]{revtex4-2}

\usepackage{amsmath,amssymb,bm}
\usepackage{graphicx}
\usepackage{booktabs}
\usepackage{hyperref}
\usepackage{comment}
\usepackage{xcolor}
\hypersetup{hidelinks}
\usepackage{adjustbox}
\usepackage{placeins}
\graphicspath{{figures/}}

\newcommand{\eq}[1]{Eq.~(\ref{#1})}
\newcommand{\eqs}[2]{Eqs.~(\ref{#1}) and (\ref{#2})}

\newcommand{\fref}[1]{FIG.\ref{#1}}
\newcommand{\sref}[1]{Section \ref{#1}}

\begin{document}

\title{Stochastic Bubble Completion in de Sitter Space: Power Spectrum, Bispectrum, and Infrared Moment Hierarchies}

\author{Hyukjung Kim}
\email{curi951007@gmail.com}
\affiliation{Lunit Inc., Gangnam-daero, Gangnam-gu, Seoul 06241, Republic of Korea. }

\author{Alp Kağan Özmen}
\email{alpozmen@std.iyte.edu.tr}
\affiliation{Department of Physics, Izmir Institute of Technology, Gülbahçe, Urla 35430, Izmir, Türkiye.}
\affiliation{Dipartimento di Fisica e Astronomia “Galileo Galilei”, Università di Padova, via Marzolo 8, 35131 Padua, Italy.}

\author{Heeseung Zoe}
\email{heeseungzoe@iyte.edu.tr}
\affiliation{Department of Physics, Izmir Institute of Technology,
Gülbahçe, Urla 35430, Izmir, Türkiye.}

\date{\today}

\begin{abstract}
Stochastic bubble nucleation makes the local completion time of a first-order phase transition a random spatial field. We study its power spectrum and equilateral bispectrum in Minkowski and de Sitter backgrounds, with particular emphasis on the infrared consequences of a finite transition duration. In an eternally extended de Sitter history, rare bubbles nucleated in the remote past generate an ancient-bubble tail that produces a hierarchy of infrared-sensitive spatial moments. For the two-point function, the first finite-$k$ correction becomes tail sensitive at $H/\beta=1/5$, while the leading moment becomes singular at $H/\beta=1/3$. The corresponding three-point thresholds are $H/\beta=1/8$ and $1/6$, showing that the bispectrum is more infrared sensitive than the power spectrum. For any finite starting time, however, the ancient-bubble tail is cut off, all spatial moments remain finite, and the strict infrared spectra recover the analytic white-noise limits ${\cal P}_\tau\propto k^3$ and ${\cal B}^{\rm eq}_\tau\propto k^6$. The eternal-background thresholds instead control the duration dependence of the moment coefficients and the crossover into an intermediate tail-sensitive regime, in agreement with our numerical calculations. We further show that temporal localization of the nucleation history provides a smooth regulator of the ancient-bubble tail and restores the same analytic infrared limits. These results establish a general first-arrival framework for completion-time fluctuations and clarify how de Sitter expansion can generate distinct power-spectrum and bispectrum infrared hierarchies relevant to curvature perturbations and non-Gaussianity.
\end{abstract}

\maketitle

\newpage

\section{Introduction}
\label{sec:intro}

First-order cosmological phase transitions provide a compelling window into early-universe physics, with stochastic gravitational waves among their best-studied possible signatures \cite{Caprini:2019egz,Athron:2023xlk}. Bubble nucleation is intrinsically stochastic, so the local completion time of the transition---the time at which a given spatial point is first reached by the true vacuum---is itself a random field. Spatial fluctuations of the transition history can in turn source curvature perturbations and have been studied in several cosmological settings \cite{Liu:2022lvz,Elor:2023xbz,Greene:2026gnw}. Bubble-induced perturbations during inflation have also been investigated \cite{Firouzjahi:2017ssb}, while transitions extending from inflation through reheating can generate isocurvature perturbations with a non-Gaussian origin \cite{Buckley:2024nen}. Related asynchronous first-order transitions during reheating have been connected to curvature perturbations and induced gravitational waves \cite{Sui:2025epg}. First-order phase transitions occurring during inflation and their dynamical evolution have been studied in explicit models \cite{An:2023jxf,Zou:2026wzi}. Beyond the purely exponential approximation, a quadratic correction in the exponent of the nucleation rate has been studied in gravitational-wave calculations, including an explicit supersymmetric flaton transition ending thermal inflation \cite{Jinno:2017ixd}, and more recently in realistic strongly supercooled models where the correction can affect density-fluctuation, primordial-black-hole, and gravitational-wave predictions \cite{Lewicki:2024sfw}. The emphasis here is different: our primary object is the spatial statistics of the stochastic completion surface itself, in particular the hierarchy of spatial moments generated by de Sitter expansion.

In the non-expanding Minkowski limit, which we use as the flat-space limit, the completion-time correlators decrease sufficiently rapidly at large separation that all spatial moments are finite. Their Fourier transforms are therefore analytic at $k=0$, giving the dimensionless infrared (IR) scalings $\mathcal P_\tau\propto k^3$ and $\mathcal B_\tau^{\rm eq}\propto k^6$. In a de Sitter background, by contrast, a bubble nucleated earlier can reach a much larger comoving region at late times, reflecting the familiar interplay between bubble growth and exponential expansion \cite{Guth:1980zm,Guth:1982pn,Turner:1992tz}. If the nucleation history is extended arbitrarily far into the past, these increasingly early and large bubbles generate a power-law spatial tail in the completion-time correlators, which we refer to below as the ancient-bubble tail. Successive spatial moments can then diverge, removing the corresponding terms from the analytic small-$k$ expansion.

This divergence is a property of the idealized nucleation history extended arbitrarily far into the past. A physical transition instead begins at a finite time, which cuts off the ancient-bubble tail and restores analyticity in the strict $k\rightarrow0$ limit. Our finite-duration analysis will show, however, that the effects of de Sitter expansion remain encoded in the growth of spatial moments with duration and in the associated crossover scales. In particular, the bispectrum moments become sensitive to the ancient-bubble tail at smaller values of $h\equiv H/\beta$ than the corresponding power-spectrum moments. This hierarchy leads to a separation between the scales at which the power spectrum and bispectrum enter their tail-dominated regimes, making the reduced non-Gaussianity sensitive to whether the two crossovers occur together or at distinct scales.

In this paper, we develop the first-arrival formalism, derive the universal power-law part of the ancient-bubble tail analytically, and characterize the three-point logarithmic enhancement through numerical evaluation of the full triple-overlap kernel. We then determine how the resulting spatial-moment hierarchy is regulated by a finite starting time and compare its duration dependence with numerical calculations. We do not assume a universal ordering of the power- and bispectrum crossover scales, and their relative location is treated as a numerical property of the finite-start problem. Whenever one spectrum has entered a tail-dominated regime while the other remains in its analytic infrared regime, the reduced equilateral non-Gaussianity develops a characteristic scale dependence. We also use a successfully completing vacuum-dominated transition as a simple illustration of the conversion $\zeta=H\delta t_c$, and we show how temporal localization of the nucleation rate suppresses the ancient-bubble tail.

The paper is organized as follows. In \sref{sec:formalism}, we define the stochastic first-arrival formalism and the scaling variables. In \sref{sec:flat}, we establish the flat-space limit, derive the one-point completion-time statistics, and determine the infrared power spectrum, bispectrum, and reduced non-Gaussianity against which the effects of expansion are compared. In \sref{sec:desitter}, we derive the eternal de Sitter tails, the finite-duration moment hierarchy, and the crossover structure. In \sref{sec:supercooled}, we discuss smooth temporal localization of the nucleation history as a dynamical infrared regulator. In \sref{sec:conclusions}, we summarize the main results. The Appendix derives the ancient-bubble asymptotics in more detail and collects the de Sitter normalization formulas.

\section{Stochastic completion-time formalism}
\label{sec:formalism}

To quantify the spatial fluctuations of the transition completion time, we employ a stochastic first-arrival formalism based on Poisson nucleation statistics, closely related to finite-bubble descriptions of completion-time fluctuations~\cite{Elor:2023xbz}. We parameterize the nucleation rate per unit physical four-volume as
\begin{equation}
\Gamma(t)=\Gamma_f\exp\left(\beta t-\frac12\gamma^2t^2\right),
\end{equation}
where $\Gamma_f$ sets the overall amplitude. After measuring time in units of the intrinsic transition scale $\beta^{-1}$, the interplay between the nucleation history and de Sitter expansion is characterized by
\begin{equation}
h\equiv\frac{H}{\beta},
\qquad
g\equiv\frac{\gamma^2}{\beta^2}.
\end{equation}
Here, $h$ measures the importance of the expansion rate relative to the transition rate, while $g$ controls the temporal localization of the nucleation history. The limit $g=0$ corresponds to an exactly exponential rate. Microscopically, semiclassical false-vacuum decay at zero temperature is described by the Coleman--Callan formalism \cite{Coleman:1977py,Callan:1977pt}, while finite-temperature nucleation is described by its thermal extension \cite{Linde:1980tt}; gravitational effects on vacuum decay are incorporated in the Coleman--De Luccia treatment \cite{Coleman:1980aw}. In the present work, however, we treat $\Gamma(t)$ as a prescribed effective time-dependent rate and assume a constant bubble-wall speed $v_w$.

For a given observation event $(x_i,t_i)$, let $V_i$ denote its past bubble-influence region, namely the spacetime region from which a bubble nucleated earlier can expand with wall speed $v_w$ and reach $x_i$ before time $t_i$. For $v_w=1$ this region coincides with the corresponding past light-cone. Assuming that nucleation events form a spatially homogeneous Poisson point process with time-dependent intensity $\Gamma(t)$, the survival probability that the local completion time $t_c(x_i)$ occurs after $t_i$ is the probability that no nucleation occurs within $V_i$. This is the same Poisson-survival construction underlying the standard false-vacuum fraction and percolation analyses of cosmological first-order transitions \cite{Guth:1982pn,Turner:1992tz},
\begin{equation}
S_i=\mathrm{Pr}[t_c(x_i)>t_i]=e^{-I_i},
\qquad
I_i=\int_{V_i}d^4x\,\sqrt{-\det g_{\mu\nu}}\,\Gamma(t).
\end{equation}
Here, $I_i$ is the mean number of nucleation events whose bubbles can reach the observation event by time $t_i$. We use the centered completion-time field
\begin{equation}
\tau(\mathbf{x})\equiv t_c(\mathbf{x})-\langle t_c\rangle,
\end{equation}
so that all connected correlators below describe fluctuations about the mean completion time.

We take Poisson nucleation as the statistical starting point and use it to construct the multi-point first-arrival statistics of the completion-time field in an expanding background. This provides the basis for deriving the power spectrum, bispectrum, and the de Sitter spatial-moment hierarchy discussed below. Spatial correlations among the nucleation sites themselves have been studied separately using stochastic and lattice approaches~\cite{DeLuca:2021mlh,Pirvu:2021roq}. These studies address an additional layer of statistics beyond the Poisson baseline considered here. A natural future extension of the present framework would therefore be to incorporate such correlated nucleation into the first-arrival construction and investigate how it modifies the completion-time correlators and their de Sitter moment hierarchy.

When evaluating correlations between multiple spatial points, their past bubble-influence regions can overlap. A bubble nucleated within an overlap region can reach more than one point and thereby correlate their completion times. The joint survival probability is therefore determined by the nucleation intensity over the union of the corresponding influence regions. For two regions, the inclusion-exclusion principle gives
\begin{equation}
I(V_1\cup V_2)=I_1+I_2-J_{12},
\end{equation}
where $J_{12}$ is the nucleation intensity integrated over the overlap $V_1\cap V_2$. For three regions,
\begin{equation}
I(V_1\cup V_2\cup V_3)
=
I_1+I_2+I_3-J_{12}-J_{13}-J_{23}+J_{123},
\end{equation}
where $J_{123}$ denotes the intensity over the common overlap $V_1\cap V_2\cap V_3$. The pairwise overlaps are subtracted to remove double counting, while the triple overlap is added back because it has been subtracted more than once. We therefore define
\begin{equation}
J_{ij}
=
\int_{V_i\cap V_j}
d^4x\sqrt{-\det g_{\mu\nu}}\,\Gamma(t),
\qquad
J_{123}
=
\int_{V_1\cap V_2\cap V_3}
d^4x\sqrt{-\det g_{\mu\nu}}\,\Gamma(t).
\end{equation}

These overlap intensities directly determine the connected completion-time correlators. Removing the disconnected statistical pieces through the cumulant expansion, the two-point and three-point spatial correlators are
\begin{equation}
C_2(r)
=
\int dt_1dt_2\,
e^{-I_1-I_2}
\left(
e^{J_{12}}-1
\right),
\end{equation}
and
\begin{equation}
C_3(r,s,\mu)
=
\int dt_1dt_2dt_3\,
K_3,
\end{equation}
where $r$ and $s$ are spatial separations, $\mu=\hat{\mathbf r}\cdot\hat{\mathbf s}$, and
\begin{equation}
K_3
=
e^{-I_1-I_2-I_3}
\left(
e^{J_{12}+J_{23}+J_{31}-J_{123}}
-e^{J_{12}}
-e^{J_{23}}
-e^{J_{31}}
+2
\right).
\end{equation}

To connect these real-space statistics with cosmological observables, we define
\begin{align}
\langle\tau_{\mathbf k}\tau_{\mathbf k'}\rangle
&=
(2\pi)^3
\delta^{(3)}(\mathbf k+\mathbf k')
P_\tau(k),
\\
\langle\tau_{\mathbf k_1}\tau_{\mathbf k_2}\tau_{\mathbf k_3}\rangle_c
&=
(2\pi)^3
\delta^{(3)}(\mathbf k_1+\mathbf k_2+\mathbf k_3)
B_\tau(k_1,k_2,k_3),
\end{align}
where
\begin{align}
P_\tau(k)
&=
\int d^3r\,
e^{-i\mathbf k\cdot\mathbf r}
C_2(r),
\\
B_\tau(k_1,k_2,k_3)
&=
\int d^3r\,d^3s\,
e^{-i\mathbf k_1\cdot\mathbf r-i\mathbf k_2\cdot\mathbf s}
C_3(r,s,\mu),
\end{align}
and their dimensionless forms are
\begin{equation}
\mathcal P_\tau(k)
=
\frac{k^3}{2\pi^2}P_\tau(k),
\qquad
\mathcal B_\tau^{\rm eq}(k)
=
\frac{k^6}{(2\pi^2)^2}B_\tau(k,k,k).
\end{equation}

The natural physical scale of the transition is the comoving bubble separation, defined through the integrated number density of nucleated bubbles,
\begin{equation}
n_b^{\rm com}
=
\int dt\,
a^3(t)\Gamma(t)e^{-I(t)},
\qquad
d_b
=
(n_b^{\rm com})^{-1/3}.
\end{equation}
Using this characteristic scale, we define
\begin{equation}
\xi\equiv kd_b,
\qquad
y\equiv\frac{kv_w}{\beta}.
\end{equation}
When logarithmic infrared corrections are present we use
\begin{equation}
L(k)
\equiv
\ln\left(\frac{k_0}{k}\right)
=
\ln\left(\frac{y_0}{y}\right),
\label{eq:Ldef}
\end{equation}
where $k_0$, equivalently $y_0=k_0v_w/\beta$, is a nonuniversal matching scale between the asymptotic infrared form and the finite-$k$ spectrum. The integer power of $L$ is universal within the asymptotic scaling discussed below, while the numerical value of $k_0$ is not.

\begin{figure*}[t]
\centering
\includegraphics[width=1\textwidth]{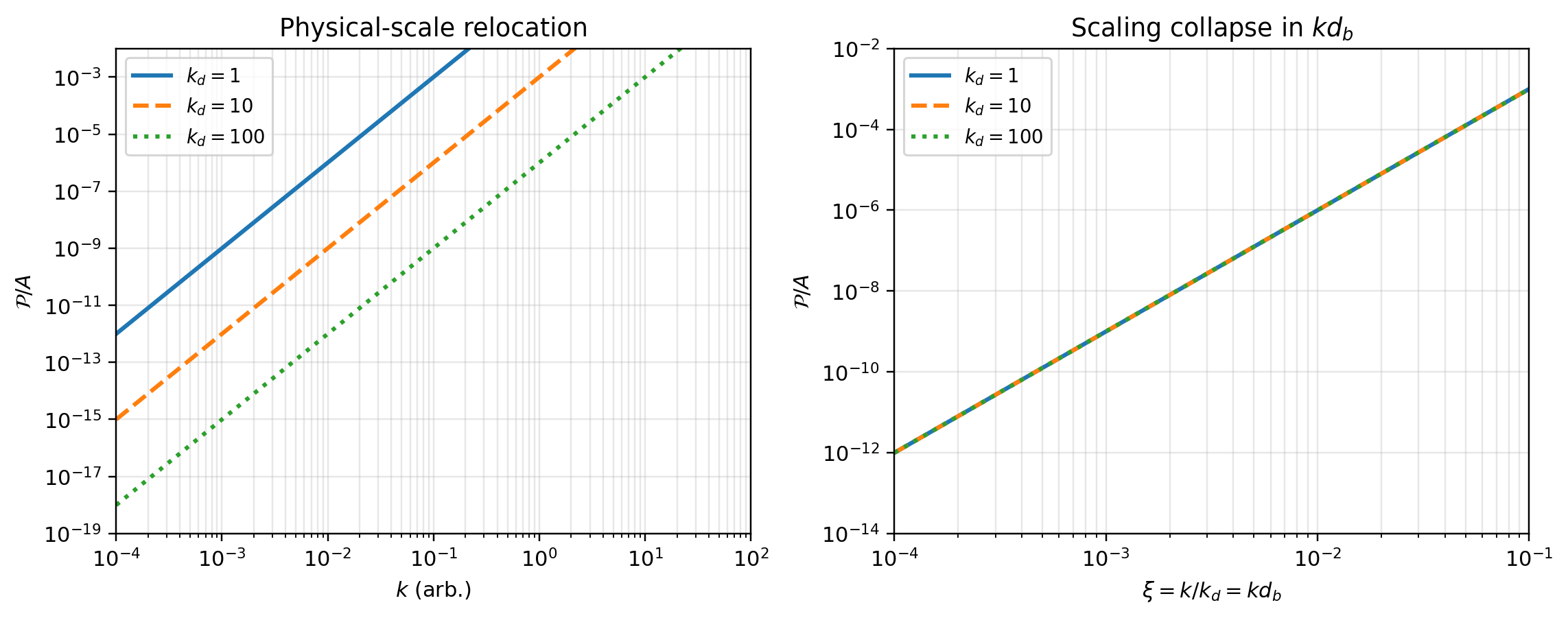}
\caption{One-scale self-similarity of the bubble statistics. Left: changing the characteristic inverse bubble separation $k_d=d_b^{-1}$ shifts the same infrared shape along the physical $k$ axis. Right: the curves collapse when expressed in the dimensionless variable $\xi=kd_b$.}
\label{fig:collapse}
\end{figure*}

For a finite-start calculation it is useful to write
\begin{equation}
\mathcal P_\tau(k)
=
\beta^{-2}{\cal F}_2(\xi;h,g,N_{\rm dS}),
\qquad
\mathcal B_\tau^{\rm eq}(k)
=
\beta^{-3}{\cal F}_3(\xi;h,g,N_{\rm dS}),
\label{eq:spectrum_observable}
\end{equation}
where $N_{\rm dS}$ denotes the duration parameter introduced below. In the eternal-background limit, or in subcritical regimes where the moments have saturated, this additional argument can be suppressed. Changing the overall transition scale shifts the feature in physical wavenumber, while plotting against $\xi=kd_b$ collapses the dimensionless shape when the remaining dimensionless parameters are fixed, as explicitly demonstrated in \fref{fig:collapse}. This self-similarity refers to the scaling structure of the nucleation geometry and should not be confused with cosmological scale invariance of the power spectrum. In the ordinary analytic infrared regime, the dimensionless completion-time spectrum instead scales as $\mathcal P_\tau\propto k^3$.

The completion-time field can be related to the curvature perturbation through a background-dependent conversion. At linear order we write
\begin{equation}
\zeta(\bm x)
=
A\,\tau(\bm x),
\label{eq:linconv}
\end{equation}
where $A$ is the conversion coefficient relating a local shift in the completion time to the corresponding change in the integrated expansion. In the $\delta N$ formalism, $A=\partial N/\partial t_c$ evaluated on the homogeneous background. For a vacuum-dominated transition with approximately constant $H$ followed by a sudden conversion onto a common post-transition hypersurface, this reduces to $A\simeq H_c$. More generally, the conversion of phase-transition inhomogeneities into curvature perturbations requires a gauge-consistent treatment of the coupled matter and metric perturbations. Recent covariant and gauge-invariant analyses of slow first-order transitions show that gauge effects can materially change the inferred curvature perturbations and their primordial-black-hole and scalar-induced-gravitational-wave consequences \cite{Franciolini:2025ztf,Wang:2026zvz}. The simple relation \eq{eq:linconv} is therefore used here only within the restricted sudden-conversion setup specified above.

The curvature spectra then inherit the completion-time correlations as
\begin{equation}
\mathcal P_\zeta
=
A^2\mathcal P_\tau,
\qquad
\mathcal B_\zeta^{\rm eq}
=
A^3\mathcal B_\tau^{\rm eq}.
\end{equation}
We define the reduced equilateral non-Gaussianity by
\begin{equation}
f_{\rm NL}^{\rm eq}(k)
\equiv
\frac{5}{18}
\frac{\mathcal B_\zeta^{\rm eq}(k)}
{\mathcal P_\zeta^2(k)}.
\label{eq:fnldef}
\end{equation}
This quantity measures the equilateral bispectrum amplitude relative to the squared power spectrum and provides a useful diagnostic of its scale dependence. The $f_{\rm NL}^{\rm eq}$ defined here should not be identified directly with the observational equilateral-template amplitude. Observational estimators use the full momentum dependence $B_\zeta(k_1,k_2,k_3)$ and project it onto a specified template, whereas \eq{eq:fnldef} is constructed only from the equilateral configuration $k_1=k_2=k_3=k$. Throughout this paper, $f_{\rm NL}^{\rm eq}$ therefore denotes this reduced equilateral diagnostic.

Whenever the connected real-space correlators possess finite spatial moments, their Fourier transforms are analytic at $k=0$. In flat spacetime the correlators decay rapidly enough for this condition to hold even for an exponentially extended nucleation history, while a finite onset in an expanding background provides an explicit causal cutoff. The dimensionless spectra therefore obey
\begin{equation}
\mathcal P_\tau
\propto
k^3,
\qquad
\mathcal B_\tau^{\rm eq}
\propto
k^6,
\label{eq:white}
\end{equation}
at sufficiently small $k$, and the reduced equilateral non-Gaussianity approaches a constant. The de Sitter effect studied below is the ancient-bubble power-law tail generated by an exponentially extended past history and the way a physical regulator reorganizes the approach to this analytic limit.

\section{Flat-space case}
\label{sec:flat}

In Minkowski spacetime, $H=0$. Without temporal localization $g=0$, the nucleation rate is exponential $\Gamma(t)=\Gamma_f e^{\beta t}$. Integrating over the flat-space bubble volume $V(t-t')=(4\pi/3)v_w^3(t-t')^3$ and setting $v_w=1$ in the explicit case gives
\begin{equation}
I(t)
=
\frac{8\pi\Gamma_f}{\beta^4}e^{\beta t}.
\end{equation}
The normalization $I(0)=1$ therefore fixes $\Gamma_f=\beta^4/(8\pi)$. The one-point completion-time distribution is
\begin{equation}
F(t)
=
\Pr[t_c\le t]
=
1-\exp\left[-e^{\beta t}\right].
\end{equation}
For the centered field $\tau=t_c-\langle t_c\rangle$, the exact centered moments are
\begin{equation}
\langle\beta\tau\rangle
=
0,
\qquad
\beta^2\langle\tau^2\rangle
=
\frac{\pi^2}{6},
\qquad
\beta^3\langle\tau^3\rangle
=
-2\zeta(3).
\end{equation}

Since the flat-space connected correlators have finite spatial moments, \eq{eq:white} fixes the leading infrared behavior to $\mathcal P_\tau\propto k^3$ and $\mathcal B_\tau^{\rm eq}\propto k^6$. To determine their amplitudes, we evaluate the connected two- and three-point completion-time correlators obtained from the joint Poisson survival probabilities, Fourier transform them, and expand the resulting expressions in the infrared variable $\xi\equiv kd_b\ll1$. The remaining real-space integrals are evaluated by deterministic numerical integration, so the coefficients below are semi-analytic predictions of the first-arrival formalism rather than fits to simulation data. For the flat-space exponential case, we obtain
\begin{align}
\beta^2\mathcal P_\tau(k)
&=
2.76136\,\xi^3
+
\mathcal O(\xi^5),
\\
\beta^3\mathcal B_\tau^{\rm eq}(k)
&=
-45.04751\,\xi^6
+
\mathcal O(\xi^8).
\end{align}
Substituting these coefficients into \eq{eq:fnldef} gives
\begin{equation}
\frac{A}{\beta}f_{\rm NL}^{\rm eq}
=
\frac{5}{18}
\frac{-45.04751}{(2.76136)^2}
=
-1.64105.
\end{equation}
Thus the flat-space exponential case has the expected analytic $k^3$ and $k^6$ infrared scalings together with a scale-independent reduced equilateral non-Gaussianity in the strict infrared.

\begin{figure*}[t]
\centering
\includegraphics[width=1\textwidth]{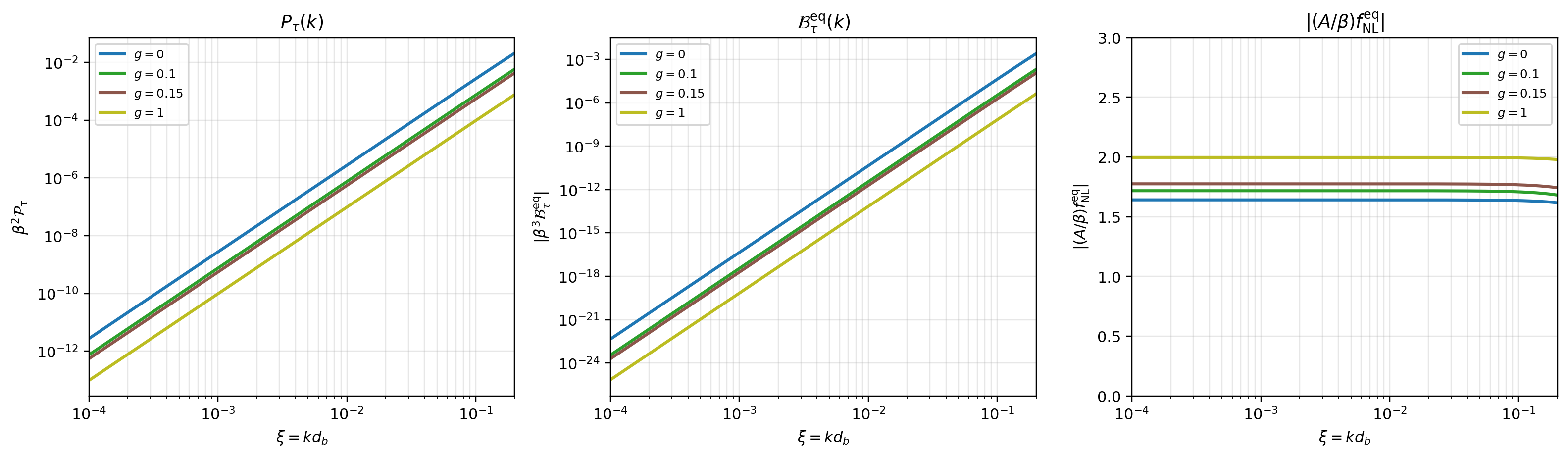}
\caption{Flat-space completion-time observables. Left: the dimensionless power spectrum follows the robust $\xi^3$ law for the displayed localization parameters $g$. Right: the intrinsic normalized combination $|(A/\beta)f_{\rm NL}^{\rm eq}|$ is scale independent in the IR. Increasing $g$ suppresses the raw power and bispectrum amplitudes much more strongly than it changes the normalized non-Gaussianity.}
\label{fig:flatmain}
\end{figure*}

For $g>0$, temporal localization suppresses the absolute amplitudes of both the power spectrum and bispectrum, with the bispectrum showing a particularly strong reduction as $g$ increases. By contrast, the normalized non-Gaussianity changes much more mildly and remains of order unity, indicating that temporal localization primarily reduces the overall fluctuation amplitudes rather than eliminating their intrinsic non-Gaussian character. Table~\ref{tab:flat} summarizes representative IR coefficients, while \fref{fig:flatmain} shows the corresponding power spectrum, bispectrum, and reduced non-Gaussianity.

\begin{table*}[t]
\begin{ruledtabular}
\begin{tabular}{cccc}
$g$ & $\beta^2\mathcal P_\tau/\xi^3$ & $\beta^3\mathcal B_\tau^{\rm eq}/\xi^6$ & $(A/\beta)f_{\rm NL}^{\rm eq}$ \\
\hline
0    & 2.76136 & -45.04751 & -1.64105 \\
0.10 & 0.75551 & -3.52898  & -1.71736 \\
0.15 & 0.55243 & -1.95019  & -1.77511 \\
1.00 & 0.09790 & -0.06888  & -1.99629 \\
\end{tabular}
\end{ruledtabular}
\caption{Flat-space IR observables for selected $g=\gamma^2/\beta^2$. The second and third columns are the coefficients multiplying $\xi^3$ in $\beta^2\mathcal P_\tau$ and $\xi^6$ in $\beta^3\mathcal B_\tau^{\rm eq}$, respectively.}
\label{tab:flat}
\end{table*}

\section{de Sitter expansion and finite-duration infrared scaling}
\label{sec:desitter}

\subsection{Power-law tails and breakdown of standard scaling in eternal de Sitter space}

To isolate the genuine dynamical effects of cosmic expansion, we first consider the idealized case of an eternally extended de Sitter background, where bubble nucleation proceeds from $t_i\to-\infty$ in an expanding metric with scale factor $a(t)=e^{Ht}$. A bubble nucleated at coordinate time $t_n$ expands at the speed of light $(v_w=1)$, reaching a late-time comoving radius
\begin{equation}
R(t_n)
=
\int_{t_n}^{\infty}
\frac{dt'}{a(t')}
=
\frac{1}{H}e^{-Ht_n}.
\end{equation}
Because the past history is unbounded, bubbles nucleated in the deep past can grow to arbitrarily large comoving sizes. We refer to these rare, giant bubbles as \emph{ancient bubbles}. A single ancient bubble can correlate widely separated points, and the resulting long-distance power-law behavior of the completion-time correlators will be referred to as the \emph{ancient-bubble tail}.

Consider two spatial points separated by a large comoving distance $r$. In flat spacetime the probability of sufficiently early nucleation is exponentially suppressed with the required bubble size, so the completion-time correlator decays rapidly and all spatial moments remain finite. In de Sitter space, by contrast, the exponential relation between nucleation time and comoving bubble radius converts the remote past into a power-law spatial tail. A non-vanishing long-distance correlation can arise when both points are engulfed by the same ancient bubble. Requiring a bubble radius of order $r$ gives
\begin{equation}
t_n
\simeq
-\frac{1}{H}\ln(Hr)
\end{equation}
up to an additive constant of order $H^{-1}$. Since $\Gamma(t_n)\propto e^{\beta t_n}$, the remote-past nucleation weight contributes the universal factor $(Hr)^{-\beta/H}=(Hr)^{-1/h}$.

Expanding the exact pair kernel at large separation, where $J_{12}\ll1$, gives $e^{J_{12}}-1=J_{12}+\mathcal O(J_{12}^2)$. The explicit dilation analysis in Appendix~\ref{app:analytic} then gives
\begin{equation}
\beta^2 C_2(r)
\sim
(Hr)^{-p}\ln^2(Hr),
\qquad
p\equiv\frac{1}{h}.
\label{eq:C2_tail}
\end{equation}
The power $(Hr)^{-p}$ follows directly from the de Sitter dilation of the nucleation rate, while the logarithmic factor arises from the completion-time integrations.

The three-point function is governed by the same power-law mechanism. We separate the overall comoving size $\lambda$ from the dimensionless triangle shape by writing
\begin{equation}
\mathbf r
=
\lambda\bar{\mathbf r},
\qquad
\mathbf s
=
\lambda\bar{\mathbf s},
\end{equation}
with $\bar{\mathbf r}$ and $\bar{\mathbf s}$ fixed as $\lambda$ is varied. The exact connected kernel satisfies
\begin{equation}
K_3
=
-e^{-I_1-I_2-I_3}J_{123}
+
\mathcal O(J^2),
\end{equation}
because the terms linear in the pairwise overlaps cancel. The dilation argument therefore fixes the leading power $(H\lambda)^{-1/h}$ independently of the detailed triangle shape.

The logarithmic degree is more subtle because the triple-overlap kernel retains nontrivial dependence on the relative completion times. Numerical evaluation of the full triple-overlap geometry is consistent, over the accessible asymptotic range, with
\begin{equation}
\beta^3
C_3
\left(
\lambda\bar{\mathbf r},
\lambda\bar{\mathbf s}
\right)
\sim
-
A_3
\left(
\bar{\mathbf r},
\bar{\mathbf s};
h
\right)
(H\lambda)^{-p}
\ln^4(H\lambda),
\qquad
p=\frac{1}{h}.
\label{eq:C3_tail}
\end{equation}
Thus the power-law exponent and the associated moment thresholds follow analytically from de Sitter dilation, while the quartic logarithmic enhancement is supported by the full-kernel numerical asymptotics. Its momentum-space consequences are also consistent with the finite-$k$ bispectrum behavior discussed below. The bispectrum thresholds derived below assume that the leading equilateral shape projection of $A_3$ does not vanish.

The physical implications of these power-law tails become transparent in momentum space. For isotropic configurations,
\begin{equation}
P_\tau(k)
=
4\pi
\int_0^\infty
dr\,r^2
\frac{\sin(kr)}{kr}
C_2(r)
=
M_0-\frac{1}{6}M_2k^2+\mathcal O(k^4),
\end{equation}
where
\begin{equation}
M_{2n}
\equiv
4\pi
\int_0^\infty
dr\,r^{2+2n}C_2(r).
\label{eq:moments}
\end{equation}
Multiplying by $k^3/(2\pi^2)$ gives the analytic infrared expansion
\begin{equation}
\mathcal P_\tau(\xi)
=
P_3\xi^3
\left[
1-\beta_2\xi^2+\mathcal O(\xi^4)
\right],
\label{eq:P_analytic}
\end{equation}
where $P_3\propto M_0$ and $\beta_2\propto M_2/M_0$.

In flat space, the exponential decay of $C_2(r)$ ensures that all moments $M_{2n}$ are finite. In eternal de Sitter space, the ancient-bubble tail
\begin{equation}
C_2(r)
\propto
(Hr)^{-1/h}
\ln^2(Hr)
\end{equation}
gives a large-distance moment integrand proportional to $r^{2+2n-1/h}\ln^2(Hr)$. Convergence therefore requires
\begin{equation}
2+2n-\frac{1}{h}<-1,
\end{equation}
which gives
\begin{equation}
h<h_c^{(n)}
\equiv
\frac{1}{3+2n}.
\end{equation}
Thus $M_0$, $M_2$, and $M_4$ remain finite for $h<1/3$, $h<1/5$, and $h<1/7$, respectively. This hierarchy determines which terms survive in the analytic small-$k$ expansion.

For the leading term, convergence of $M_0$ for $h<1/3$ preserves the standard white-noise behavior $\mathcal P_\tau\propto\xi^3$. When $h>1/3$, $M_0$ diverges in the eternal-background problem and the leading behavior becomes nonanalytic. The first analytic correction requires convergence of $M_2$, which holds only for $h<1/5$. Thus the successive spatial moments become sensitive to the ancient-bubble tail before the leading zero mode itself becomes divergent.

The bispectrum exhibits the same mechanism, but the two independent separation vectors produce a six-dimensional spatial measure. In the equilateral configuration,
\begin{equation}
B_\tau(k,k,k)
=
\int
d^3\mathbf r\,d^3\mathbf s\,
e^{-i(\mathbf k_1\cdot\mathbf r+\mathbf k_2\cdot\mathbf s)}
C_3(\mathbf r,\mathbf s).
\end{equation}
Expanding in the long-wavelength limit and performing the angular averages gives
\begin{equation}
B_\tau(k,k,k)
=
B_0
+
k^2\widetilde B_2
+
\mathcal O(k^4),
\end{equation}
with
\begin{align}
B_0
&\equiv
\int
d^3\mathbf r\,d^3\mathbf s\,
C_3(\mathbf r,\mathbf s),
\\
\widetilde B_2
&\equiv
\int
d^3\mathbf r\,d^3\mathbf s
\left[
-\frac{r^2+s^2}{6}
+
\frac{\mathbf r\cdot\mathbf s}{6}
\right]
C_3(\mathbf r,\mathbf s).
\end{align}
The sign of the mixed term follows from $\mathbf k_1\cdot\mathbf k_2=-k^2/2$ in the equilateral configuration. Multiplying by $(k^3/2\pi^2)^2$ gives
\begin{equation}
\mathcal B_\tau^{\rm eq}(\xi)
=
B_6\xi^6
\left[
1-\alpha_2\xi^2+\mathcal O(\xi^4)
\right],
\end{equation}
where $B_6\propto B_0$ and
\begin{equation}
\alpha_2
=
-\frac{\widetilde B_2}{d_b^2B_0}.
\end{equation}

Writing the three-point configuration in terms of the overall comoving scale $\lambda$ and dimensionless shape variables gives
\begin{equation}
d^3\mathbf r\,d^3\mathbf s
\propto
\lambda^5d\lambda\,d\mu_{\rm sh}.
\end{equation}
Using \eq{eq:C3_tail}, the $2n$-th order bispectrum moment therefore has the large-$\lambda$ radial behavior
\begin{equation}
\lambda^{5+2n-1/h}
\ln^4(H\lambda).
\end{equation}
Convergence requires
\begin{equation}
5+2n-\frac{1}{h}<-1,
\end{equation}
giving
\begin{equation}
h<h_{c,B}^{(n)}
\equiv
\frac{1}{6+2n}.
\end{equation}
Comparing this with the power-spectrum condition $h<1/(3+2n)$ gives the combined hierarchy summarized in Table~\ref{tab:critical_rates}.

\begin{table}[h]
\centering
\begin{tabular}{lcc}
\hline\hline
Observable & Quadratic correction ($n=1$) & Leading term ($n=0$) \\
\hline
Power spectrum ($\mathcal P_\tau$) & $h_c^{(1)}=1/5$ & $h_c^{(0)}=1/3$ \\
Equilateral bispectrum ($\mathcal B_\tau^{\rm eq}$) & $h_{c,B}^{(1)}=1/8$ & $h_{c,B}^{(0)}=1/6$ \\
\hline\hline
\end{tabular}
\caption{Critical expansion rates above which the corresponding spatial moments diverge in the idealized eternal de Sitter limit. For any finite-duration transition, these eternal-background thresholds instead determine the duration sensitivity of the strict infrared coefficients and govern the intermediate tail behavior.}
\label{tab:critical_rates}
\end{table}

The bispectrum therefore becomes sensitive to the ancient-bubble tail at smaller $h$ than the power spectrum. This statement concerns moment convergence as $h$ is varied and does not by itself determine the ordering of the finite-duration crossover wavenumbers at fixed $h$.

\subsection{Finite-duration transitions and causal restoration of standard scaling}

The nonanalytic scalings and divergent spatial moments obtained in the eternal de Sitter limit arise from extending the nucleation history to $t_i\to-\infty$. In realistic cosmological settings, such as a first-order transition terminating thermal inflation~\cite{Lyth:1995hj,Lyth:1995ka,Easther:2008sx,Kim:2026mal} or a first-order nonthermal transition after inflation in which expanding bubbles nucleate during the preheating era~\cite{Khlebnikov:1998sz}, nucleation begins at a finite time $t_i$.

A finite onset time imposes a causal upper bound on the comoving reach of bubbles. For de Sitter expansion,
\begin{equation}
R_{\rm H}(t_i)
\equiv
v_w
\int_{t_i}^{\infty}
\frac{dt'}{a(t')}
=
\frac{v_w}{H}e^{-Ht_i}.
\label{eq:Rmax}
\end{equation}
If the transition ends at a finite reference time $t_f$, the actual maximum radius is
\begin{equation}
R_{\max}(t_i,t_f)
=
\frac{v_w}{H}
\left(
e^{-Ht_i}-e^{-Ht_f}
\right),
\end{equation}
which has the same $e^{H|t_i|}$ scaling for a sufficiently long transition. We denote the corresponding large-distance cutoff by $R_{\rm cut}\sim R_{\rm H}$, since order-unity geometric factors do not affect the critical exponents.

The finite cutoff makes the spatial moments explicit functions of the past duration. For the power spectrum,
\begin{equation}
M_{2n}(t_i)
\equiv
4\pi
\int_0^{R_{\rm cut}(t_i)}
dr\,
r^{2+2n}
C_2(r).
\label{eq:M_reg}
\end{equation}
For the bispectrum, let $\mathcal W_{2n}(\mathbf r,\mathbf s)$ denote the homogeneous degree-$2n$ polynomial multiplying the $k^{2n}$ term in the equilateral Fourier expansion and define
\begin{equation}
\mathcal M_{B,2n}(t_i)
\equiv
\int_{\Omega(\mathbf r,\mathbf s)\le R_{\rm cut}(t_i)}
d^3\mathbf r\,d^3\mathbf s\,
\mathcal W_{2n}(\mathbf r,\mathbf s)
C_3(\mathbf r,\mathbf s),
\label{eq:B_reg}
\end{equation}
where
\begin{equation}
\Omega(\mathbf r,\mathbf s)
\equiv
\max
\left(
r,s,|\mathbf r-\mathbf s|
\right)
\end{equation}
provides an overall measure of the three-point configuration size up to order-unity geometric factors. The leading weight is $\mathcal W_0=1$, while
\begin{equation}
\mathcal W_2(\mathbf r,\mathbf s)
=
-\frac{r^2+s^2}{6}
+
\frac{\mathbf r\cdot\mathbf s}{6}
\end{equation}
reproduces $\widetilde B_2$.

The duration dependence is determined by the eternal-background thresholds. Below the corresponding threshold the moment approaches a finite asymptotic value as the past duration increases, while above it the ancient-bubble contribution becomes increasingly dominated by the causal cutoff. Suppressing duration-independent dimensional factors, the cutoff-dependent contributions scale as
\begin{align}
M_{2n}(t_i)
&\propto
(HR_{\rm cut})^{3+2n-1/h}
\ln^2(HR_{\rm cut}),
\\
\mathcal M_{B,2n}(t_i)
&\propto
(HR_{\rm cut})^{6+2n-1/h}
\ln^4(HR_{\rm cut}).
\end{align}
For a sufficiently long de Sitter interval,
\begin{equation}
HR_{\rm cut}
\propto
e^{N_{\rm dS}},
\qquad
N_{\rm dS}
\equiv
H|t_i|,
\end{equation}
so that
\begin{equation}
M_{2n}(t_i)
\propto
e^{(3+2n-1/h)N_{\rm dS}}
N_{\rm dS}^2,
\qquad
\mathcal M_{B,2n}(t_i)
\propto
e^{(6+2n-1/h)N_{\rm dS}}
N_{\rm dS}^4.
\label{eq:moment_duration_scaling}
\end{equation}
At the corresponding critical values the power-law factor becomes marginal and the overall-scale integration generates one additional power of $N_{\rm dS}$.

For any finite $t_i$, the cutoff keeps all spatial moments finite and restores analyticity at $k=0$. On scales much larger than the maximum bubble size,
\begin{equation}
\xi
\ll
\xi_{\rm cut}
\equiv
\frac{d_b}{R_{\rm cut}(t_i)}
\sim
Hd_b\,e^{Ht_i},
\label{eq:xi_cut_def}
\end{equation}
where the last relation assumes $v_w\simeq1$, the dimensionless spectra recover
\begin{align}
\mathcal P_\tau(\xi)
&=
P_3(t_i)\xi^3
\left[
1-\beta_2(t_i)\xi^2+\mathcal O(\xi^4)
\right],
\\
\mathcal B_\tau^{\rm eq}(\xi)
&=
B_6(t_i)\xi^6
\left[
1-\alpha_2(t_i)\xi^2+\mathcal O(\xi^4)
\right].
\end{align}
The coefficients remain duration dependent, with $P_3(t_i)\propto M_0(t_i)$ and $B_6(t_i)\propto\mathcal M_{B,0}(t_i)$.

The same ancient-bubble contribution can nevertheless remain relevant on scales smaller than the causal cutoff scale, allowing the tail-sensitive behavior to persist over an intermediate momentum range with $\xi\gtrsim\xi_{\rm cut}$. For a sufficiently long nucleation history, the spectra can therefore cross from the analytic deep-infrared regime to behavior inherited from the eternal-background ancient-bubble tail.

\begin{figure}[h]
\centering
\includegraphics[width=\textwidth]{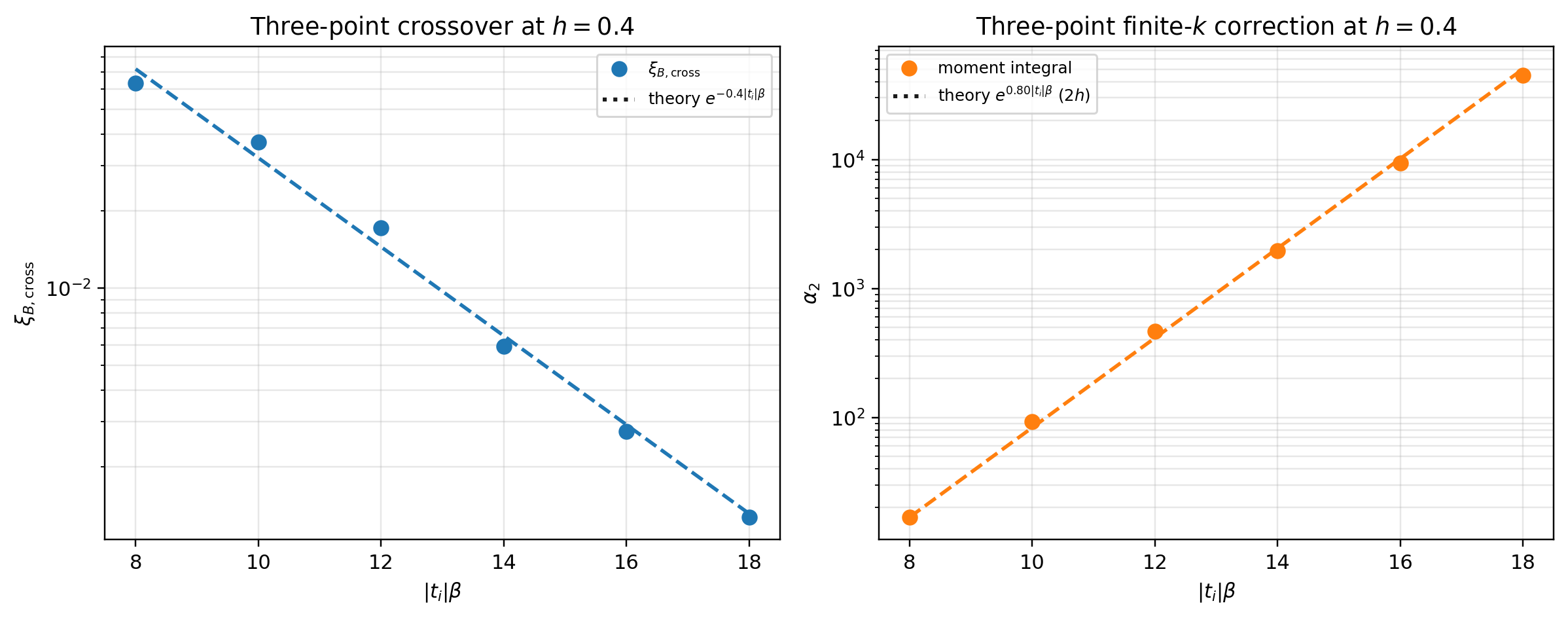}
\caption{Finite-duration scaling of the three-point sector in the $h>1/6$ branch, illustrated at $h=0.40$. Left: the bispectrum moment scale $\xi_{B,2}=\alpha_2^{-1/2}$ as a function of $|t_i|\beta$. Right: the quadratic coefficient $\alpha_2$. The numerical points are obtained from the finite-start moment calculation, and the fitted duration dependence is compared with the $h>1/6$ branch of \eq{eq:alpha2_cases}.}
\label{fig:finite-start-scaling}
\end{figure}

\begin{figure}[t]
\centering
\includegraphics[width=\textwidth]{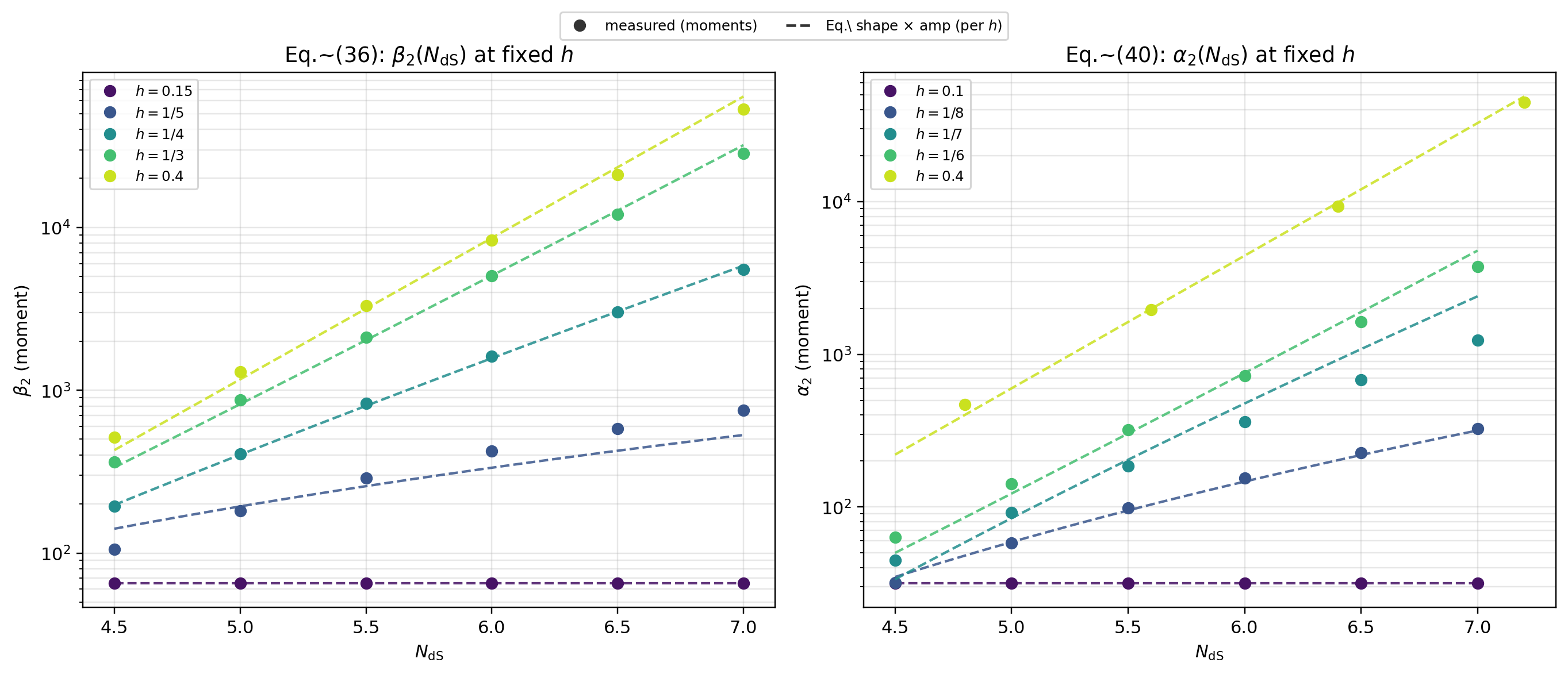}
\caption{Duration dependence of the power-spectrum coefficient $\beta_2$ and bispectrum coefficient $\alpha_2$ for representative values of $h$. The numerical points are obtained from the finite-start moment ratios in \eqs{eq:beta2_def}{eq:alpha2_def}, while the dashed curves show fits to the corresponding asymptotic branches in \eqs{eq:beta2_cases}{eq:alpha2_cases}.}
\label{fig:moment-hierarchy}
\end{figure}

To study this duration dependence, we vary the initial nucleation time $t_i$ while keeping $h$, $g$, and $v_w$ fixed. For each $t_i$, the time origin is chosen by imposing $I(0)=1$, which fixes $\Gamma_f$, and the mean bubble separation $d_b$ is recomputed from the corresponding finite-start bubble density before forming $\xi=kd_b$. The resulting duration scans are shown in \fref{fig:finite-start-scaling} and \fref{fig:moment-hierarchy}.

We distinguish three related but conceptually different scales. The causal scale $\xi_{\rm cut}=d_b/R_{\rm cut}$ marks the boundary of the deep-infrared regime in which wavelengths exceed the maximum causal bubble reach and the spectra are necessarily analytic. The moment scales
\begin{equation}
\xi_{P,2}
\equiv
\beta_2^{-1/2},
\qquad
\xi_{B,2}
\equiv
\alpha_2^{-1/2},
\label{eq:moment_cross_scales}
\end{equation}
indicate where the first analytic $k^2$ corrections become order unity. Finally, $\xi_{P,\rm tail}$ and $\xi_{B,\rm tail}$ characterize the crossover of the leading spectral behavior itself. They are defined operationally as the intersections of the fitted finite-start analytic and tail forms over the fitting windows used in the corresponding figures. Their numerical values therefore carry a fitting-window dependence and should be interpreted as crossover diagnostics rather than sharply defined physical scales. The causal, moment, and tail scales probe different aspects of the finite-duration problem and need not coincide.

For the power spectrum,
\begin{equation}
\beta_2
=
\frac{M_2}{6d_b^2M_0}.
\label{eq:beta2_def}
\end{equation}
Using $C_2(r)\propto(Hr)^{-p}\ln^2(Hr)$ together with $HR_{\rm cut}\propto e^{N_{\rm dS}}$ gives
\begin{equation}
\beta_2
\propto
\begin{cases}
\text{const.}, & h<1/5,\\
N_{\rm dS}^3, & h=1/5,\\
e^{(5-1/h)N_{\rm dS}}N_{\rm dS}^2, & 1/5<h<1/3,\\
e^{2N_{\rm dS}}/N_{\rm dS}, & h=1/3,\\
e^{2N_{\rm dS}}, & h>1/3.
\end{cases}
\label{eq:beta2_cases}
\end{equation}
The corresponding moment scale is $\xi_{P,2}=\beta_2^{-1/2}$.

To evaluate the equilateral moments $B_0$ and $\widetilde B_2$, we parameterize the two separation lengths as
\begin{equation}
r=\lambda,
\qquad
s=\lambda q,
\qquad
0\le q\le1,
\end{equation}
where $\lambda$ is the dimensionful overall comoving scale introduced above. Denoting by $C_{3\ell}(\lambda,\lambda q)$ the Legendre components of the angular dependence of $C_3$, the moments can be written as
\begin{align}
B_0
&=
\int d\lambda\,
\lambda^5
\int_0^1dq\,
q^2
C_{30}(\lambda,\lambda q),
\label{eq:B0_proj}
\\
\widetilde B_2
&=
\int d\lambda\,
\lambda^7
\int_0^1dq\,
q^2
\left[
-\frac{1+q^2}{6}C_{30}(\lambda,\lambda q)
+
\frac{q}{6}C_{31}(\lambda,\lambda q)
\right].
\label{eq:B2_proj}
\end{align}
Here $C_{30}$ gives the angle-independent contribution, while $C_{31}$ arises from the mixed term proportional to $\mathbf r\cdot\mathbf s$. The quadratic correction coefficient is
\begin{equation}
\alpha_2
=
-\frac{\widetilde B_2}{d_b^2B_0}.
\label{eq:alpha2_def}
\end{equation}

Using
\begin{equation}
C_3(\lambda,\lambda q)
\propto
-(H\lambda)^{-p}
\ln^4(H\lambda)
\end{equation}
for fixed $q$, and assuming that the leading equilateral shape projection does not vanish, the duration dependence becomes
\begin{equation}
\alpha_2
\propto
\begin{cases}
\text{const.}, & h<1/8,\\
N_{\rm dS}^5, & h=1/8,\\
e^{(8-1/h)N_{\rm dS}}N_{\rm dS}^4, & 1/8<h<1/6,\\
e^{2N_{\rm dS}}/N_{\rm dS}, & h=1/6,\\
e^{2N_{\rm dS}}, & h>1/6.
\end{cases}
\label{eq:alpha2_cases}
\end{equation}
The corresponding bispectrum moment scale is $\xi_{B,2}=\alpha_2^{-1/2}$.

The numerical points in \fref{fig:finite-start-scaling} and \fref{fig:moment-hierarchy} are obtained by applying the finite-start prescription above, evaluating the corresponding moment ratios for each $t_i$, and constructing $\beta_2$, $\alpha_2$, $\xi_{P,2}$, and $\xi_{B,2}$. Their duration dependence is then compared with the asymptotic predictions in \eqs{eq:beta2_cases}{eq:alpha2_cases}. \fref{fig:finite-start-scaling} focuses on the three-point sector for the supercritical example $h=0.40$, while \fref{fig:moment-hierarchy} compares the power- and bispectrum coefficients over several values of $h$.

The choice $h=0.40$ in \fref{fig:finite-start-scaling} is made to provide a clear numerical test of the supercritical duration scaling. It is used here as a diagnostic example rather than as a representative parameter choice for a successfully completing vacuum-dominated transition. For such a slow transition, with $\beta/H=2.5$, completion and percolation must be assessed separately because cosmic expansion can compete with bubble nucleation and growth~\cite{Guth:1982pn,Turner:1992tz}.

\subsection{Hierarchical crossovers and scale-dependent non-Gaussianity}
\label{subsec:crossovers}

For $\xi\gtrsim\xi_{\rm cut}$, one has $k^{-1}\lesssim R_{\rm cut}$, so the mode probes spatial separations that are not yet limited by the finite-start cutoff. In this intermediate regime, the momentum-space spectra can inherit the scaling associated with the ancient-bubble tail. The power spectrum obeys
\begin{equation}
\mathcal P_\tau(k)
\propto
\begin{cases}
k^3, & h<1/3,\\
k^3L^3(k), & h=1/3,\\
k^{1/h}L^2(k), & h>1/3,
\end{cases}
\label{eq:P_formal}
\end{equation}
while
\begin{equation}
\mathcal B_\tau^{\rm eq}(k)
\propto
\begin{cases}
k^6, & h<1/6,\\
k^6L^5(k), & h=1/6,\\
k^{1/h}L^4(k), & h>1/6.
\end{cases}
\label{eq:B_formal}
\end{equation}
For every finite starting time these relations describe, at most, an intermediate tail-dominated regime. At sufficiently small $k$, finite spatial moments restore $\mathcal P_\tau\propto k^3$ and $\mathcal B_\tau^{\rm eq}\propto k^6$.

The corresponding formal tail scaling of the reduced equilateral non-Gaussianity is
\begin{equation}
f_{\rm NL}^{\rm eq}(k)
\propto
\begin{cases}
\text{const.}, & h<1/6,\\
L^5(k), & h=1/6,\\
k^{1/h-6}L^4(k), & 1/6<h<1/3,\\
k^{-3}L^{-2}(k), & h=1/3,\\
k^{-1/h}, & h>1/3.
\end{cases}
\label{eq:fnl_formal}
\end{equation}
This displays directly that the bispectrum becomes tail sensitive at a smaller expansion rate than the leading power spectrum.

For a finite-duration transition the scale-dependent running is instead controlled by $\xi_{P,\rm tail}$ and $\xi_{B,\rm tail}$. Their ordering is not fixed by the critical values alone. If the bispectrum enters its tail-dominated regime first,
\begin{equation}
\xi_{B,\rm tail}
\ll
\xi
\ll
\xi_{P,\rm tail},
\end{equation}
then
\begin{equation}
f_{\rm NL}^{\rm eq}
\propto
k^{1/h-6}L^4(k).
\label{eq:fnl_mixed_B}
\end{equation}
If instead the power spectrum crosses first,
\begin{equation}
\xi_{P,\rm tail}
\ll
\xi
\ll
\xi_{B,\rm tail},
\end{equation}
then
\begin{equation}
f_{\rm NL}^{\rm eq}
\propto
k^{6-2/h}L^{-4}(k).
\label{eq:fnl_mixed_P}
\end{equation}
Thus the existence and width of a mixed regime are numerical properties of the finite-duration problem, whereas the exponents follow from the real-space tail hierarchy.

\begin{figure*}[htbp!]
\centering
\rotatebox{90}{%
  \begin{minipage}{0.9\textheight}
    \centering
    \includegraphics[width=\linewidth]{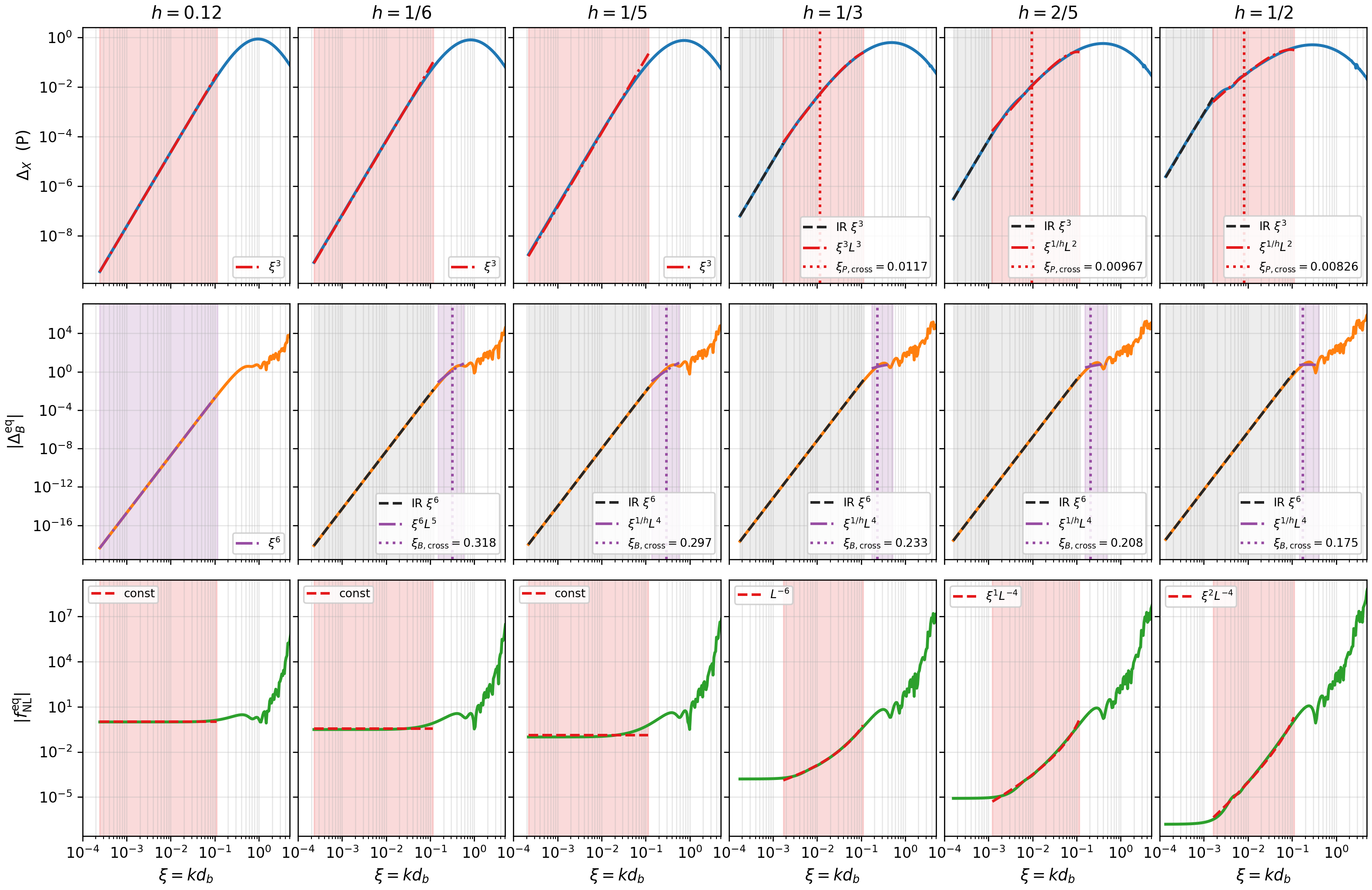}
    \caption{Power spectrum (top), equilateral bispectrum (middle), and reduced equilateral non-Gaussianity (bottom) for representative expansion rates $h$. The deep-IR region is compared with the analytic finite-duration forms, while intermediate fitting windows are compared with \eqs{eq:P_formal}{eq:B_formal}. The relative leading-tail crossover locations determine which, if either, of \eqs{eq:fnl_mixed_B}{eq:fnl_mixed_P} is realized.}
    \label{fig:matched-crossovers}
  \end{minipage}%
}
\end{figure*}

\fref{fig:matched-crossovers} shows the numerical power spectrum, equilateral bispectrum, and reduced non-Gaussianity for representative values of $h$. The shaded intervals are finite fitting windows over which the corresponding analytic or tail form describes the numerical spectrum. They should not be interpreted as extrapolations of the anomalous laws to $k\to0$, and the extracted crossover locations retain some dependence on the chosen fitting windows.

\FloatBarrier

\subsection{Curvature perturbations in vacuum-dominated transitions}
\label{subsec:vacuum_conversion}

To connect the completion-time statistics with curvature perturbations, we consider a minimal vacuum-dominated transition that successfully completes and reheats suddenly onto a common final-density hypersurface. In the separate-universe or $\delta N$ formalism \cite{Sasaki:1995aw,Lyth:2005fi}, a local shift of the completion time changes the integrated expansion between an initial flat hypersurface and the final uniform-density hypersurface. If the Hubble rate is approximately constant during the relevant interval,
\begin{equation}
\zeta(\mathbf x)
=
\delta N(\mathbf x)
=
H\,\delta t_c(\mathbf x)
=
H\tau(\mathbf x).
\label{eq:zeta_tau_linear}
\end{equation}
Within this case the conversion is linear, $N_{,t_c}=H$ and $N_{,t_ct_c}=0$, so there is no additional conversion-generated bispectrum. The identification of $\tau$ with a curvature perturbation requires the transition to complete and the locally reheated regions to admit a common post-transition uniform-density hypersurface. It therefore does not apply directly to the original nonpercolating old-inflation regime. Local first-arrival statistics may still be defined there, but they cannot in general be identified with a global primordial curvature spectrum through $\mathcal P_\zeta=H^2\mathcal P_\tau$. The example below should instead be understood as a successfully completing vacuum-dominated first-order transition \cite{Guth:1980zm,Guth:1982pn,Turner:1992tz}.

For a fast transition, we take $g=0$ and $h=H/\beta\ll1$. In this regime $h$ lies well below the critical values associated with the power-spectrum and bispectrum moments, so no parametrically separated tail-dominated regime develops before the bubble scale is reached. The canonical $k^3$ and $k^6$ infrared scalings therefore persist over the relevant infrared range. The curvature spectra are
\begin{equation}
\mathcal P_\zeta^{\rm bub}
=
H^2\mathcal P_\tau
\propto
h^2\xi^3,
\qquad
\mathcal B_\zeta^{\rm bub,eq}
=
H^3\mathcal B_\tau^{\rm eq}
\propto
h^3\xi^6.
\end{equation}
Using the flat exponential coefficients, the intrinsic reduced non-Gaussianity is
\begin{equation}
f_{\rm NL}^{\rm bubble}
\simeq
-\frac{1.641}{h}.
\end{equation}
Thus the bubble contribution can have a large intrinsic reduced non-Gaussianity even when its power is strongly suppressed on scales $\xi\ll1$.

\begin{figure}[h]
\centering
\includegraphics[width=\textwidth]{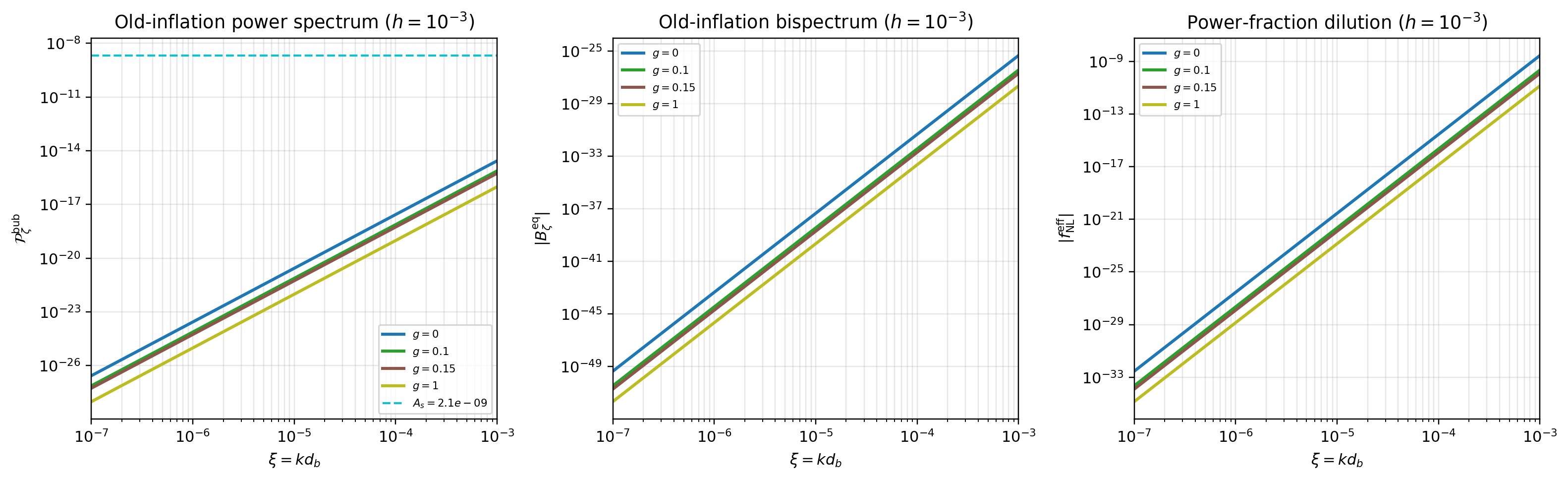}
\caption{Bubble-induced curvature observables for a fast, successfully completing vacuum-dominated transition with $h=10^{-3}$, well below all of the moment thresholds discussed above. Left: $\mathcal P_\zeta^{\rm bub}$. Middle: $|\mathcal B_\zeta^{\rm bub,eq}|$. Right: the reduced non-Gaussianity after dilution by the total scalar power. The plot illustrates the curvature conversion for this deeply subcritical example. Comparison with an observed comoving scale additionally requires specifying the physical value of $d_b$.}
\label{fig:vacuum-conversion}
\end{figure}

If an independent nearly Gaussian component dominates the observed scalar power,
\begin{equation}
\mathcal P_\zeta^{\rm tot}
=
\mathcal P_\zeta^{\rm G}
+
\mathcal P_\zeta^{\rm bub},
\end{equation}
while the bispectrum remains dominated by the bubble contribution, then
\begin{equation}
f_{\rm NL}^{\rm eff}
=
f_{\rm NL}^{\rm bubble}
\left(
\frac{\mathcal P_\zeta^{\rm bub}}
{\mathcal P_\zeta^{\rm tot}}
\right)^2.
\end{equation}
Here $f_{\rm NL}^{\rm eff}$ is again only a reduced equilateral amplitude, and a direct CMB constraint would require the full bispectrum shape and template projection. \fref{fig:vacuum-conversion} illustrates this conversion for a representative fast transition. A direct comparison with the CMB amplitude $A_s\simeq2.1\times10^{-9}$~\cite{Planck:2018vyg} additionally requires specifying the physical mapping between $d_b$ and the observed comoving wavenumber. The figure should therefore be interpreted as an illustration of the curvature conversion rather than as a model-independent observational constraint.

\section{Smooth temporal localization as a dynamical infrared regulator}
\label{sec:supercooled}

The previous section showed that a finite starting time $t_i$ cuts off the ancient-bubble tail for $g=0$. A different regularization can arise from the time dependence of the nucleation rate itself. The usual exponential approximation follows from linearizing the nucleation action around a characteristic time, giving $\Gamma(t)\propto e^{\beta t}$. Beyond this approximation, a quadratic term in the exponent produces a Gaussian-like temporal localization of the nucleation history. Such corrections have been considered in studies of strongly supercooled transitions~\cite{Megevand:2016lpr,Athron:2022mmm,Athron:2023xlk} and in calculations of gravitational-wave and density-fluctuation observables~\cite{Jinno:2017ixd,Lewicki:2024sfw}. In~\cite{Jinno:2017ixd}, the quadratic correction was also evaluated for an explicit supersymmetric flaton transition ending thermal inflation, while \cite{Lewicki:2024sfw} showed that the second-order term can be important in realistic strongly supercooled models. Motivated by this structure, we use
\begin{equation}
\Gamma(t)
=
\Gamma_f
\exp\left(
\beta t-\frac12\gamma^2t^2
\right),
\qquad
g
=
\frac{\gamma^2}{\beta^2}>0,
\end{equation}
as a minimal effective parametrization. Our normalization of $\gamma$ matches that of \cite{Lewicki:2024sfw}. In~\cite{Jinno:2017ixd}, the quadratic term is written as $-\gamma_J^2t^2$, corresponding to $\gamma=\sqrt{2}\gamma_J$ in our convention. Neglecting the slowly varying prefactor, our parametrization corresponds to expanding the tunneling action as $S(t)=S_0-\beta t+\tfrac12\gamma^2t^2+\cdots$. The parameter $g$ therefore measures the temporal localization of the nucleation history, while its relation to the underlying microscopic potential remains model dependent.

The Gaussian profile requires some care when extrapolated to arbitrarily late times in exact de Sitter space. For a comoving worldline observed at $t\to\infty$, the integrated Poisson intensity approaches
\begin{equation}
I_\infty
=
\frac{4\pi v_w^3}{3H^3}
\int_{-\infty}^{\infty}
dt\,\Gamma(t),
\label{eq:Iinfty_gaussian}
\end{equation}
which is finite for the Gaussian-localized rate. Hence $S_\infty=e^{-I_\infty}>0$, so this profile by itself does not guarantee completion of every worldline. We therefore use the $g>0$ profile to describe the suppression of nucleation in the remote past, with the late-time evolution understood either to be matched onto a history that completes the transition or to be restricted to a finite active interval with a specified completion prescription. The Gaussian profile therefore serves here as an effective description of the remote-past suppression of ancient bubbles rather than as a complete global transition history.

Using $t_n\simeq-H^{-1}\ln(Hr)$ at large separation gives
\begin{equation}
\Gamma(r)
\propto
(Hr)^{-1/h}
\exp\left[
-\frac{g}{2h^2}\ln^2(Hr)
\right].
\label{eq:gamma_dynamical_cutoff}
\end{equation}
Compared with the pure ancient-bubble power-law tail obtained for $g=0$, the additional log-normal factor suppresses the contribution from increasingly early and large bubbles more strongly than any fixed power of $r$. Consequently, all spatial moments associated with the ancient-bubble tail remain finite for $g>0$. With a completing late-time prescription, the strict $k\to0$ power spectrum and bispectrum therefore retain their analytic $k^3$ and $k^6$ behavior.

For sufficiently small $g$, however, the log-normal suppression need not become important immediately. Over an intermediate range of scales, the spectra can still retain the scaling associated with the ancient-bubble tail before crossing back to their analytic infrared forms. A matching estimate gives
\begin{align}
\mathcal P_\tau(k)
&\propto
k^{1/h}L^2(k)
\exp\left[
-\frac{g}{2h^2}L^2(k)
\right],
\label{eq:P_g_dependence}
\\
\mathcal B_\tau^{\rm eq}(k)
&\propto
k^{1/h}L^4(k)
\exp\left[
-\frac{g}{2h^2}L^2(k)
\right].
\label{eq:B_g_dependence}
\end{align}
These expressions describe only this intermediate regime. They must eventually cross over to $\mathcal P_\tau\propto k^3$ and $\mathcal B_\tau^{\rm eq}\propto k^6$ as $k\to0$.

If there is a range of scales in which the power spectrum follows \eq{eq:P_g_dependence} while the bispectrum remains in its analytic $k^6$ regime, the reduced equilateral non-Gaussianity behaves as
\begin{equation}
f_{\rm NL}^{\rm eq}(k)
\propto
k^{6-2/h}L^{-4}(k)
\exp\left[
+\frac{g}{h^2}L^2(k)
\right].
\label{eq:fnl_modified_running}
\end{equation}
The positive exponential does not imply a divergence as $k\to0$, because \eq{eq:fnl_modified_running} applies only within the intermediate regime and ceases to hold once the power spectrum returns to its analytic infrared behavior.

\begin{figure*}[htbp!]
\centering
\rotatebox{90}{%
  \begin{minipage}{0.9\textheight}
    \centering
    \includegraphics[width=\linewidth]{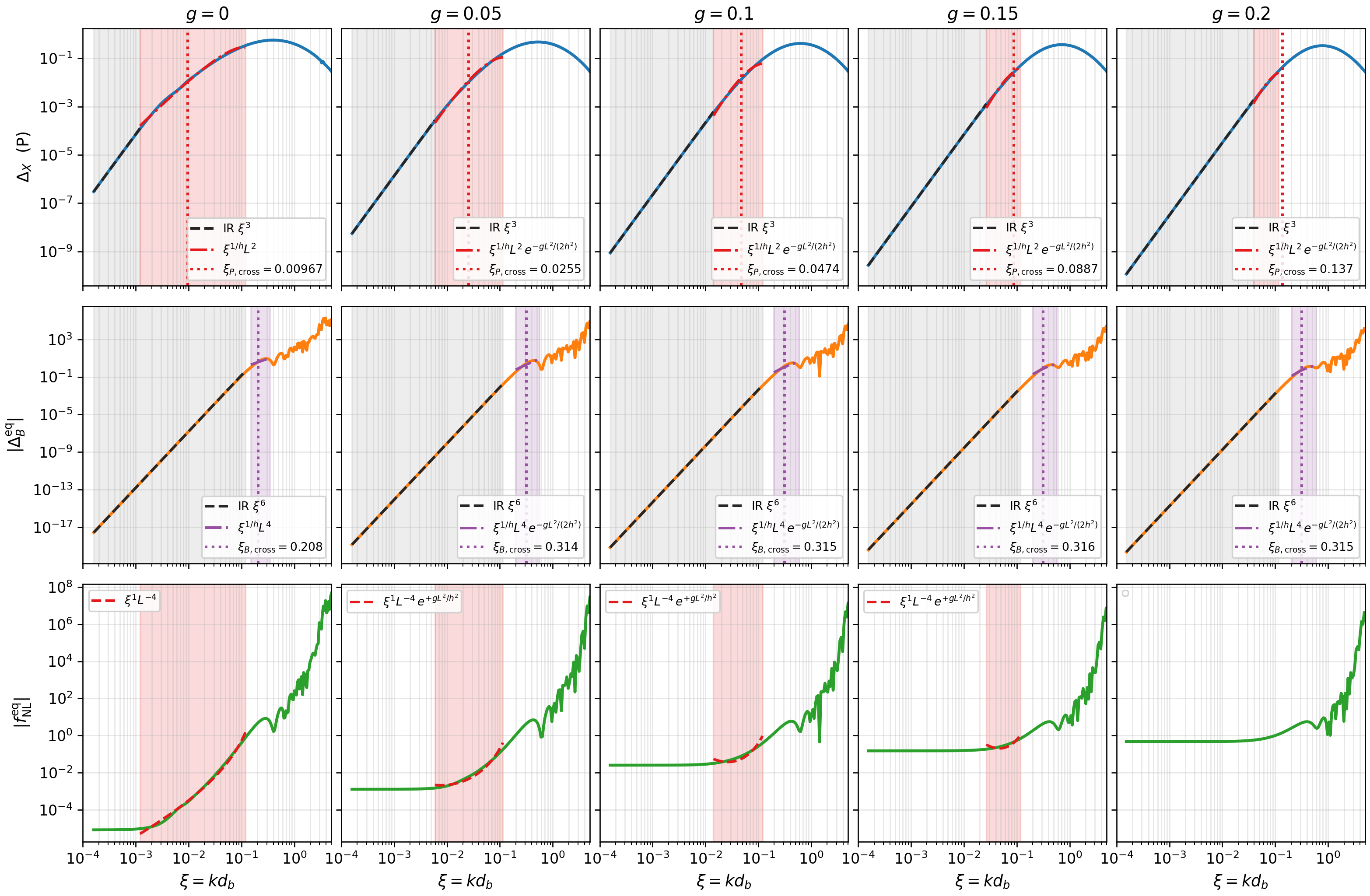}
    \caption{Power spectrum (top), equilateral bispectrum (middle), and reduced equilateral non-Gaussianity (bottom) at fixed $h=0.4$ for representative values of $g$. Shaded intervals indicate the finite $\xi$ ranges over which the analytic infrared forms or the localized-tail estimates in \eqs{eq:P_g_dependence}{eq:fnl_modified_running} are compared with the numerical results. The localized-tail expressions describe only the intermediate regime and are not extrapolated to $k\to0$.}
    \label{fig:g-dependencies-crossovers}
  \end{minipage}%
}
\end{figure*}

\fref{fig:g-dependencies-crossovers} shows this tail-regulation effect numerically at fixed $h=0.4$ while varying $g$. Increasing $g$ suppresses the ancient-bubble contribution and shortens the finite momentum interval over which the tail-sensitive scaling is visible. The figure is therefore intended as a diagnostic of the smooth infrared regulation rather than as a complete model of a globally completing transition.

\FloatBarrier

\section{Discussion}
\label{sec:conclusions}

The completion-time power spectrum and bispectrum are controlled by the same ancient-bubble geometry but probe different spatial measures. In the eternally extended exponential de Sitter problem, the analytically determined power-law tail gives the power-spectrum hierarchy $h_c^{(n)}=1/(3+2n)$, while the six-dimensional three-point measure gives $h_{c,B}^{(n)}=1/(6+2n)$. The first quadratic corrections become tail sensitive at $h=1/5$ and $h=1/8$, and the leading zero modes at $h=1/3$ and $h=1/6$, respectively. The associated logarithmic enhancements are $\ln^2(Hr)$ for the two-point tail and, consistently with the full-kernel numerical asymptotics, $\ln^4(H\lambda)$ for the three-point tail.

A finite starting time changes the interpretation of these eternal-background singularities rather than erasing their physical content. All spatial moments are finite for any finite duration, so the strict IR always returns to $\mathcal P_\tau\propto k^3$ and $\mathcal B_\tau^{\rm eq}\propto k^6$. The eternal-background thresholds instead govern the duration dependence of the moment coefficients and the crossover into an intermediate tail-sensitive regime. The numerical duration scans of $\beta_2$ and $\alpha_2$ are consistent with the corresponding piecewise asymptotic laws.

The reduced non-Gaussianity is controlled by the relative leading-tail crossover scales. The critical values $1/3$ and $1/6$ do not impose a universal ordering of $\xi_{P,\rm tail}$ and $\xi_{B,\rm tail}$. If one spectrum has entered its tail-sensitive regime while the other remains analytic, $f_{\rm NL}^{\rm eq}$ follows one of the mixed scalings in \eqs{eq:fnl_mixed_B}{eq:fnl_mixed_P}. Once both are tail dominated, the formal eternal scaling approaches $f_{\rm NL}^{\rm eq}\propto k^{-1/h}$ for $h>1/3$. The crossover locations themselves are finite-duration numerical diagnostics and retain some dependence on the fitting procedure.

Temporal localization supplies a second regulator of the ancient-bubble tail. For $g>0$, the ancient-bubble weight acquires a log-normal suppression and its contribution to all spatial moments converges. A purely Gaussian rate extended to $t\to\infty$ does not by itself guarantee completion, however. The localized profile must be matched onto a completing late-time history or supplemented by an equivalent completion prescription. With that qualification, the tail formulas describe only a finite intermediate window and the spectra return to the analytic $k^3$ and $k^6$ laws in the strict IR.

For a successfully completing vacuum-dominated transition followed by sudden equal-density reheating, $\zeta=H\delta t_c$ is linear, so the curvature spectra inherit the completion-time statistics without an additional conversion-generated bispectrum. In the fast-transition case $h\ll1$, the intrinsic bubble non-Gaussianity is large, $f_{\rm NL}^{\rm bubble}\propto h^{-1}$, but the bubble power on scales much larger than the bubble separation is strongly suppressed. The original nonpercolating old-inflation regime lies outside this curvature-conversion interpretation because it does not provide the common post-transition hypersurface assumed by the $\delta N$ construction. More generally, recent covariant and gauge-invariant studies emphasize that the conversion of stochastic transition inhomogeneities into curvature perturbations is background and gauge dependent \cite{Franciolini:2025ztf,Wang:2026zvz}. Any observational interpretation of a completing transition must therefore keep the conversion prescription, the bubble power fraction, and the physical mapping of $d_b$ explicit.

Several extensions follow naturally from the present framework. A concrete application to thermal inflation would combine the bubble-completion contribution with thermally generated and processed primordial scalar sources, while determining the nucleation history from a microscopic tunneling model~\cite{Bae:2025vkn,Jinno:2017ixd,Kim:2026mal,Lewicki:2024sfw}. An important next step is to determine the full bispectrum $B_\tau(k_1,k_2,k_3)$ beyond the equilateral configuration. Together with a model-dependent conversion to the curvature bispectrum and an explicit projection onto observational bispectrum templates, this would establish the relation between the reduced equilateral diagnostic $f_{\rm NL}^{\rm eq}$ used here and observable non-Gaussianity parameters. The Poisson assumption can also be relaxed to include correlated nucleation~\cite{DeLuca:2021mlh,Pirvu:2021roq}, while a gauge-consistent evolution through the transition would provide a more complete connection between the completion-time statistics and cosmological observables~\cite{Franciolini:2025ztf,Wang:2026zvz}. These extensions would test the robustness and observational implications of the infrared moment hierarchy beyond the idealized setup considered here.

\begin{acknowledgments}
This work is supported by the TÜBİTAK ARDEB 1001 program under Project No.~123F257. The authors thank Ahmet Şekerci and Adem Kılıç for the helpful discussions.  HZ thanks Kiwoon Choi, Sang Hui Im, and Won Sang Cho for their hospitality during his visit to IBS-CTPU.
\end{acknowledgments}

\appendix

\section{Ancient-bubble asymptotics and de Sitter normalization}
\label{app:analytic}

\subsection{Ancient-bubble power counting}
\label{app:tailpower}

We first consider the exponential nucleation history with $g=0$ in the eternal de Sitter background. A constant wall speed $v_w$ is understood throughout this subsection and does not affect the dilation exponents. The power-law part of the ancient-bubble tail follows from the relation between nucleation time and comoving bubble size.

For a single observation event at $(\mathbf x_i,t_i)$,
\begin{equation}
I_i
=
\int_{-\infty}^{t_i}
dt_n\,
a^3(t_n)\Gamma(t_n)
V_i^{\rm com}(t_n;t_i),
\label{eq:app_Ii}
\end{equation}
where
\begin{equation}
V_i^{\rm com}
=
\frac{4\pi}{3}R_i^3,
\qquad
R_i(t_n)
=
\frac{v_w}{H}
\left(
e^{-Ht_n}-e^{-Ht_i}
\right).
\label{eq:app_Vi_Ri}
\end{equation}
The corresponding pair and triple overlap intensities are
\begin{equation}
J_{12}(r;t_1,t_2)
=
\int_{-\infty}^{\min(t_1,t_2)}
dt_n\,
a^3(t_n)\Gamma(t_n)
V_{12}^{\rm com}
\left[
r;R_1(t_n),R_2(t_n)
\right],
\label{eq:app_J12_overlap}
\end{equation}
and
\begin{equation}
J_{123}(\mathbf r,\mathbf s;t_1,t_2,t_3)
=
\int_{-\infty}^{\min(t_1,t_2,t_3)}
dt_n\,
a^3(t_n)\Gamma(t_n)
V_{123}^{\rm com},
\label{eq:app_J123_overlap}
\end{equation}
where $V_{12}^{\rm com}$ and $V_{123}^{\rm com}$ are the common comoving volumes of the corresponding bubble-influence spheres.

To expose the large-distance scaling, define
\begin{equation}
X_n
\equiv
\frac{e^{-Ht_n}}{H},
\qquad
X_i
\equiv
\frac{e^{-Ht_i}}{H},
\qquad
p
\equiv
\frac{\beta}{H}
=
\frac{1}{h}.
\label{eq:app_X}
\end{equation}
Then $R_i=v_w(X_n-X_i)$ and, since $HX_n=e^{-Ht_n}$,
\begin{equation}
dt_n
=
-\frac{dX_n}{HX_n},
\qquad
a^3(t_n)
=
(HX_n)^{-3},
\qquad
\Gamma(t_n)
=
\Gamma_f e^{\beta t_n}
=
\Gamma_f(HX_n)^{-p}.
\label{eq:app_X_rate}
\end{equation}

We separate the overall size of a spatial configuration from its dimensionless shape by writing
\begin{equation}
\mathbf r_{ij}
=
\lambda\,
\bar{\mathbf r}_{ij},
\qquad
u_i
\equiv
Ht_i+\ln(H\lambda),
\label{eq:app_lambda_def}
\end{equation}
where $\lambda$ has dimensions of length and the barred separations specify a fixed dimensionless shape. For the two-point function we choose $\bar r_{12}=1$, so that $\lambda=r$. For the three-point function $\lambda$ is the overall size of the triangle, while the barred vectors specify its shape. The same decomposition can therefore be used both to determine the dilation law and to parameterize the spatial integrals over configuration space.

Writing
\begin{equation}
X_i
=
\lambda e^{-u_i},
\qquad
X_n
=
\lambda z,
\label{eq:app_scaled_X}
\end{equation}
gives
\begin{equation}
dt_n
=
-\frac{dz}{Hz},
\qquad
a^3(t_n)
=
(H\lambda z)^{-3},
\qquad
\Gamma(t_n)
=
\Gamma_f(H\lambda z)^{-p},
\qquad
V_{1\cdots m}^{\rm com}
=
\lambda^3
\overline V_{1\cdots m}^{\rm com}.
\label{eq:app_scaled_factors}
\end{equation}
The factor $\lambda^{-3}$ from $a^3$ cancels the factor $\lambda^3$ from the comoving overlap volume, while $dz/z$ is invariant. The only nontrivial power of the overall scale therefore comes from the nucleation rate, giving
\begin{equation}
J_{1\cdots m}
=
(H\lambda)^{-p}
\widehat J_m
\left(
\{u_i\};
\{\bar{\mathbf r}_{ij}\};
h
\right).
\label{eq:app_J_scaling}
\end{equation}
Thus the universal ancient-bubble power $(H\lambda)^{-1/h}$ follows directly from de Sitter dilation, independently of the detailed overlap geometry.

With the normalization $I(0)=1$,
\begin{equation}
I(t_i)
=
e^{\beta t_i}
=
\exp
\left[
p
\left(
u_i-\ln(H\lambda)
\right)
\right].
\label{eq:app_I_scaled}
\end{equation}

For the two-point function, the small-overlap expansion gives
\begin{equation}
C_2(r)
=
\int dt_1dt_2\,
e^{-I_1-I_2}J_{12}
+
\mathcal O(J_{12}^2).
\label{eq:app_C2_linear}
\end{equation}
Using $\lambda=r$, \eq{eq:app_J_scaling}, and $dt_i=du_i/H$ gives
\begin{equation}
\beta^2
C_2(\lambda)
\simeq
p^2(H\lambda)^{-p}
\int du_1du_2\,
\exp
\left[
-e^{p[u_1-\ln(H\lambda)]}
-e^{p[u_2-\ln(H\lambda)]}
\right]
\widehat J_2(u_1,u_2;h).
\label{eq:app_C2_scaled_integral}
\end{equation}
The large-$H\lambda$ asymptotic evaluation of \eq{eq:app_C2_scaled_integral} gives
\begin{equation}
\beta^2
C_2(\lambda)
\sim
(H\lambda)^{-p}
\left[
a_2(h)\ln^2(H\lambda)
+
a_1(h)\ln(H\lambda)
+
a_0(h)
+
\cdots
\right],
\label{eq:app_C2_tail}
\end{equation}
and hence
\begin{equation}
\beta^2 C_2(r)
\sim
(Hr)^{-1/h}
\ln^2(Hr)
\label{eq:app_C2_leading}
\end{equation}
at leading logarithmic order.

For the connected three-point function,
\begin{equation}
C_3(\mathbf r,\mathbf s)
=
-\int dt_1dt_2dt_3\,
e^{-I_1-I_2-I_3}
J_{123}
+
\mathcal O(J^2),
\label{eq:app_C3_linear}
\end{equation}
where the terms linear in the pairwise overlaps cancel, leaving the irreducible triple overlap as the leading connected contribution. For fixed $\bar{\mathbf r}$ and $\bar{\mathbf s}$,
\begin{equation}
\beta^3
C_3
\left(
\lambda\bar{\mathbf r},
\lambda\bar{\mathbf s}
\right)
\simeq
-p^3(H\lambda)^{-p}
\int du_1du_2du_3\,
\exp
\left[
-\sum_{i=1}^3
e^{p[u_i-\ln(H\lambda)]}
\right]
\widehat J_3
\left(
u_1,u_2,u_3;
\bar{\mathbf r},
\bar{\mathbf s},
h
\right).
\label{eq:app_C3_scaled_integral}
\end{equation}

The power $(H\lambda)^{-p}$ is fixed by the dilation argument above. The logarithmic degree depends on the relative-time structure of the triple-overlap geometry and cannot be inferred by simply counting the three completion-time integrations. Numerical evaluation of the full triple-overlap kernel is consistent, over the accessible asymptotic range, with
\begin{equation}
\beta^3
C_3
\left(
\lambda\bar{\mathbf r},
\lambda\bar{\mathbf s}
\right)
\sim
-
A_3
\left(
\bar{\mathbf r},
\bar{\mathbf s};
h
\right)
(H\lambda)^{-p}
\ln^4(H\lambda)
\label{eq:app_C3_tail}
\end{equation}
for generic fixed triangle shapes. The power-law exponent is therefore fixed analytically, while the quartic logarithm is supported by the full-kernel numerical asymptotics and by the corresponding momentum-space behavior discussed in the main text.

The same variable $\lambda$ now acts as the overall-scale coordinate of configuration space. For the two-point function $r=\lambda$, so $d^3r\propto\lambda^2d\lambda$. For the three-point function the six-dimensional measure separates into an overall scale and shape variables as
\begin{equation}
d^3r\,d^3s
\propto
\lambda^5d\lambda\,d\mu_{\rm sh},
\label{eq:app_measure}
\end{equation}
up to a $\lambda$-independent shape Jacobian. The large-distance parts of the zero-mode integrals therefore behave as
\begin{equation}
\int d^3r\,C_2(r)
\propto
\int^\infty d\lambda\,
\lambda^{2-p}
\ln^2(H\lambda),
\qquad
\int d^3r\,d^3s\,
C_3(\mathbf r,\mathbf s)
\propto
-
\int^\infty d\lambda\,
\lambda^{5-p}
\ln^4(H\lambda),
\label{eq:app_zero_modes}
\end{equation}
where the second relation assumes a nonvanishing leading shape projection.

The corresponding convergence thresholds are $p=3$ and $p=6$, or equivalently $h=1/3$ and $h=1/6$. Near either threshold, introducing the dimensionless scale $x=H\lambda$ gives
\begin{equation}
\int_1^\infty
dx\,
x^{-1-\epsilon}
\ln^n x
=
\frac{\Gamma(n+1)}{\epsilon^{n+1}},
\qquad
\epsilon>0.
\label{eq:app_log_integral}
\end{equation}
Thus $n=2$ with $p=3+\epsilon$ gives a cubic pole in the power-spectrum zero-mode coefficient, while $n=4$ with $p=6+\epsilon$ gives a fifth-order pole in the corresponding bispectrum coefficient.

At the power-spectrum threshold $h=1/3$, one has $p=3$ and
\begin{equation}
C_2(r)
\propto
r^{-3}\ln^2(Hr).
\label{eq:app_C2_critical_real}
\end{equation}
If $r_*$ denotes the distance beyond which the ancient-bubble asymptotic form applies, the leading infrared part of the Fourier transform is
\begin{equation}
P_\tau(k)
=
4\pi
\int_0^\infty
dr\,r^2
C_2(r)
\frac{\sin kr}{kr}
\sim
\int_{r_*}^{1/k}
\frac{dr}{r}
\ln^2(Hr).
\label{eq:app_power_critical_transform}
\end{equation}
Defining
\begin{equation}
k_0
\equiv
\frac{1}{r_*},
\qquad
L(k)
\equiv
\ln\frac{k_0}{k},
\label{eq:app_matching_scale}
\end{equation}
one finds
\begin{equation}
\int_{r_*}^{1/k}
\frac{dr}{r}
\ln^2(Hr)
=
\frac{1}{3}L^3(k)
+
\mathcal O\!\left(L^2(k)\right),
\label{eq:app_power_log}
\end{equation}
where the lower logarithmic powers depend on the precise matching at $r\sim r_*$. Since $\mathcal P_\tau(k)\propto k^3P_\tau(k)$,
\begin{equation}
\mathcal P_\tau(k)
\propto
k^3L^3(k),
\qquad
h=\frac13.
\label{eq:app_power_critical}
\end{equation}

At the bispectrum threshold $h=1/6$, one has $p=6$. After separating the shape variables, the overall-scale part of the Fourier transform is marginal in the same way,
\begin{equation}
B_\tau^{\rm eq}(k)
\sim
\int_{\lambda_*}^{1/k}
\frac{d\lambda}{\lambda}
\ln^4(H\lambda)
\propto
L^5(k),
\label{eq:app_bispectrum_critical_transform}
\end{equation}
where the nonuniversal matching scale is absorbed into the definition of $L(k)$ and affects only lower logarithmic powers. Hence
\begin{equation}
\mathcal B_\tau^{\rm eq}(k)
\propto
k^6L^5(k),
\qquad
h=\frac16.
\label{eq:app_bispectrum_critical}
\end{equation}

Above the corresponding eternal-background thresholds, the nonanalytic ancient-bubble tails give
\begin{equation}
\mathcal P_\tau(k)
\propto
k^{1/h}L^2(k),
\qquad
h>\frac13,
\label{eq:app_power_supercritical}
\end{equation}
and
\begin{equation}
\mathcal B_\tau^{\rm eq}(k)
\propto
k^{1/h}L^4(k),
\qquad
h>\frac16.
\label{eq:app_bispectrum_supercritical}
\end{equation}
For any finite starting time the ancient-bubble tail is eventually cut off, all spatial moments remain finite, and the strict $k\to0$ spectra return to the analytic forms $\mathcal P_\tau\propto k^3$ and $\mathcal B_\tau^{\rm eq}\propto k^6$.

\subsection{De Sitter normalization}

For $g=0$ in de Sitter space and $v_w=1$, the normalization condition $I(0)=1$ gives
\begin{equation}
\Gamma_f
=
\frac{
\beta^4(1+h)(1+2h)(1+3h)
}{
8\pi
}.
\label{eq:app_Gammaf}
\end{equation}
The mean comoving bubble separation can then be written as
\begin{equation}
d_b
=
\frac{D(h)}{\beta},
\qquad
D^3(h)
=
\frac{
8\pi
}{
(1+h)(1+2h)(1+3h)\Gamma(1+3h)
}.
\label{eq:app_db}
\end{equation}

\bibliographystyle{apsrev4-2}
\bibliography{bubblesref}

@article{Caprini:2019egz,
    author = "Caprini, Chiara and others",
    title = "{Detecting gravitational waves from cosmological phase transitions with LISA: an update}",
    eprint = "1910.13125",
    archivePrefix = "arXiv",
    primaryClass = "astro-ph.CO",
    reportNumber = "DESY-19-159, IPPP/19/27, HIP-2019-14/TH, MITP/19-066, IFT-UAM/CSIC-19-139",
    doi = "10.1088/1475-7516/2020/03/024",
    journal = "JCAP",
    volume = "03",
    pages = "024",
    year = "2020"
}

@article{Athron:2023xlk,
    author = "Athron, Peter and Bal{\'a}zs, Csaba and Fowlie, Andrew and Morris, Lachlan and Wu, Lei",
    title = "{Cosmological phase transitions: From perturbative particle physics to gravitational waves}",
    eprint = "2305.02357",
    archivePrefix = "arXiv",
    primaryClass = "hep-ph",
    doi = "10.1016/j.ppnp.2023.104094",
    journal = "Prog. Part. Nucl. Phys.",
    volume = "135",
    pages = "104094",
    year = "2024"
}

@article{Liu:2022lvz,
    author = "Liu, Jing and Bian, Ligong and Cai, Rong-Gen and Guo, Zong-Kuan and Wang, Shao-Jiang",
    title = "{Constraining First-Order Phase Transitions with Curvature Perturbations}",
    eprint = "2208.14086",
    archivePrefix = "arXiv",
    primaryClass = "astro-ph.CO",
    doi = "10.1103/PhysRevLett.130.051001",
    journal = "Phys. Rev. Lett.",
    volume = "130",
    number = "5",
    pages = "051001",
    year = "2023"
}

@article{Elor:2023xbz,
    author = "Elor, Gilly and Jinno, Ryusuke and Kumar, Soubhik and McGehee, Robert and Tsai, Yuhsin",
    title = "{Finite Bubble Statistics Constrain Late Cosmological Phase Transitions}",
    eprint = "2311.16222",
    archivePrefix = "arXiv",
    primaryClass = "hep-ph",
    reportNumber = "FTPI-MINN-23-20, UTWI-39-2023",
    doi = "10.1103/PhysRevLett.133.211003",
    journal = "Phys. Rev. Lett.",
    volume = "133",
    number = "21",
    pages = "211003",
    year = "2024"
}

@misc{Greene:2026gnw,
    author = "Greene, Kylar and Ho, Daven Wei Ren and Kumar, Soubhik and Tsai, Yuhsin",
    title = "{Cosmological and Astrophysical Constraints on Late First-Order Phase Transitions}",
    eprint = "2603.00272",
    archivePrefix = "arXiv",
    primaryClass = "hep-ph",
    month = "2",
    year = "2026"
}

@article{Firouzjahi:2017ssb,
    author = "Firouzjahi, Hassan and Jazayeri, Sadra and Karami, Asieh and Rostami, Tahereh",
    title = "{Bubble nucleation and inflationary perturbations}",
    eprint = "1707.07550",
    archivePrefix = "arXiv",
    primaryClass = "gr-qc",
    doi = "10.1088/1475-7516/2017/12/029",
    journal = "JCAP",
    volume = "12",
    pages = "029",
    year = "2017"
}

@article{An:2023jxf,
    author = "An, Haipeng and Su, Boye and Tai, Hanwen and Wang, Lian-Tao and Yang, Chen",
    title = "{Phase transition during inflation and the gravitational wave signal at pulsar timing arrays}",
    eprint = "2308.00070",
    archivePrefix = "arXiv",
    primaryClass = "astro-ph.CO",
    doi = "10.1103/PhysRevD.109.L121304",
    journal = "Phys. Rev. D",
    volume = "109",
    number = "12",
    pages = "L121304",
    year = "2024"
}

@article{Zou:2026wzi,
    author = "Zou, Jintao and Bian, Ligong and Wang, Shao-Jiang",
    title = "{Simulating first-order phase transition during inflation}",
    eprint = "2602.05501",
    archivePrefix = "arXiv",
    primaryClass = "hep-ph",
    doi = "10.1103/3hz7-n5t5",
    journal = "Phys. Rev. D",
    volume = "113",
    number = "12",
    pages = "123514",
    year = "2026"
}

@article{Coleman:1977py,
    author = "Coleman, Sidney R.",
    title = "{The Fate of the False Vacuum. 1. Semiclassical Theory}",
    reportNumber = "HUTP-77-A004",
    doi = "10.1103/PhysRevD.15.2929",
    journal = "Phys. Rev. D",
    volume = "15",
    pages = "2929--2936",
    year = "1977",
    note = "[Erratum: Phys.Rev.D 16, 1248 (1977)]"
}

@article{Callan:1977pt,
    author = "Callan, Jr., Curtis G. and Coleman, Sidney R.",
    title = "{The Fate of the False Vacuum. 2. First Quantum Corrections}",
    reportNumber = "HUTP-77-A032",
    doi = "10.1103/PhysRevD.16.1762",
    journal = "Phys. Rev. D",
    volume = "16",
    pages = "1762--1768",
    year = "1977"
}

@article{Coleman:1980aw,
    author = "Coleman, Sidney R. and De Luccia, Frank",
    title = "{Gravitational Effects on and of Vacuum Decay}",
    reportNumber = "SLAC-PUB-2463",
    doi = "10.1103/PhysRevD.21.3305",
    journal = "Phys. Rev. D",
    volume = "21",
    pages = "3305",
    year = "1980"
}

@article{Guth:1980zm,
    author = "Guth, Alan H.",
    editor = "Fang, Li-Zhi and Ruffini, R.",
    title = "{The Inflationary Universe: A Possible Solution to the Horizon and Flatness Problems}",
    reportNumber = "SLAC-PUB-2576",
    doi = "10.1103/PhysRevD.23.347",
    journal = "Phys. Rev. D",
    volume = "23",
    pages = "347--356",
    year = "1981"
}

@article{Guth:1982pn,
    author = "Guth, Alan H. and Weinberg, Erick J.",
    title = "{Could the Universe Have Recovered from a Slow First Order Phase Transition?}",
    reportNumber = "MIT-CTP-950",
    doi = "10.1016/0550-3213(83)90307-3",
    journal = "Nucl. Phys. B",
    volume = "212",
    pages = "321--364",
    year = "1983"
}

@article{Turner:1992tz,
    author = "Turner, Michael S. and Weinberg, Erick J. and Widrow, Lawrence M.",
    title = "{Bubble nucleation in first order inflation and other cosmological phase transitions}",
    reportNumber = "FERMILAB-PUB-91-334-A, CU-TP-558, IASSNS-HEP-92-21",
    doi = "10.1103/PhysRevD.46.2384",
    journal = "Phys. Rev. D",
    volume = "46",
    pages = "2384--2403",
    year = "1992"
}

@article{Sasaki:1995aw,
    author = "Sasaki, Misao and Stewart, Ewan D.",
    title = "{A General analytic formula for the spectral index of the density perturbations produced during inflation}",
    eprint = "astro-ph/9507001",
    archivePrefix = "arXiv",
    reportNumber = "LANCS-TH-9504, OU-TAP-22",
    doi = "10.1143/PTP.95.71",
    journal = "Prog. Theor. Phys.",
    volume = "95",
    pages = "71--78",
    year = "1996"
}

@article{Lyth:2005fi,
    author = "Lyth, David H. and Rodriguez, Yeinzon",
    title = "{The Inflationary prediction for primordial non-Gaussianity}",
    eprint = "astro-ph/0504045",
    archivePrefix = "arXiv",
    doi = "10.1103/PhysRevLett.95.121302",
    journal = "Phys. Rev. Lett.",
    volume = "95",
    pages = "121302",
    year = "2005"
}

@article{Megevand:2016lpr,
    author = "Megevand, Ariel and Ramirez, Santiago",
    title = "{Bubble nucleation and growth in very strong cosmological phase transitions}",
    eprint = "1611.05853",
    archivePrefix = "arXiv",
    primaryClass = "astro-ph.CO",
    doi = "10.1016/j.nuclphysb.2017.03.009",
    journal = "Nucl. Phys. B",
    volume = "919",
    pages = "74--109",
    year = "2017"
}

@article{Athron:2022mmm,
    author = "Athron, Peter and Bal{\'a}zs, Csaba and Morris, Lachlan",
    title = "{Supercool subtleties of cosmological phase transitions}",
    eprint = "2212.07559",
    archivePrefix = "arXiv",
    primaryClass = "hep-ph",
    doi = "10.1088/1475-7516/2023/03/006",
    journal = "JCAP",
    volume = "03",
    pages = "006",
    year = "2023"
}

@article{Planck:2018vyg,
    author = "Aghanim, N. and others",
    collaboration = "Planck",
    title = "{Planck 2018 results. VI. Cosmological parameters}",
    eprint = "1807.06209",
    archivePrefix = "arXiv",
    primaryClass = "astro-ph.CO",
    doi = "10.1051/0004-6361/201833910",
    journal = "Astron. Astrophys.",
    volume = "641",
    pages = "A6",
    year = "2020",
    note = "[Erratum: Astron.Astrophys. 652, C4 (2021)]"
}

@article{Buckley:2024nen,
    author = "Buckley, Matthew R. and Du, Peizhi and Fernandez, Nicolas and Weikert, Mitchell J.",
    title = "{Dark radiation isocurvature from cosmological phase transitions}",
    eprint = "2402.13309",
    archivePrefix = "arXiv",
    primaryClass = "hep-ph",
    doi = "10.1088/1475-7516/2024/07/031",
    journal = "JCAP",
    volume = "07",
    pages = "031",
    year = "2024"
}

@article{DeLuca:2021mlh,
    author = "De Luca, Valerio and Franciolini, Gabriele and Riotto, Antonio",
    title = "{Bubble correlation in first-order phase transitions}",
    eprint = "2110.04229",
    archivePrefix = "arXiv",
    primaryClass = "hep-ph",
    doi = "10.1103/PhysRevD.104.123539",
    journal = "Phys. Rev. D",
    volume = "104",
    number = "12",
    pages = "123539",
    year = "2021"
}

@article{Pirvu:2021roq,
    author = "Pirvu, Dalila and Braden, Jonathan and Johnson, Matthew C.",
    title = "{Bubble clustering in cosmological first order phase transitions}",
    eprint = "2109.04496",
    archivePrefix = "arXiv",
    primaryClass = "hep-th",
    doi = "10.1103/PhysRevD.105.043510",
    journal = "Phys. Rev. D",
    volume = "105",
    number = "4",
    pages = "043510",
    year = "2022"
}

@article{Lyth:1995ka,
    author = "Lyth, David H. and Stewart, Ewan D.",
    title = "{Thermal inflation and the moduli problem}",
    eprint = "hep-ph/9510204",
    archivePrefix = "arXiv",
    reportNumber = "LANCS-TH-9505, RESCEU-16-95, LANCASTER-TH-9505",
    doi = "10.1103/PhysRevD.53.1784",
    journal = "Phys. Rev. D",
    volume = "53",
    pages = "1784--1798",
    year = "1996"
}

@article{Lyth:1995hj,
    author = "Lyth, David H. and Stewart, Ewan D.",
    title = "{Cosmology with a TeV mass GUT Higgs}",
    eprint = "hep-ph/9502417",
    archivePrefix = "arXiv",
    reportNumber = "LANCASTER-TH-9502",
    doi = "10.1103/PhysRevLett.75.201",
    journal = "Phys. Rev. Lett.",
    volume = "75",
    pages = "201--204",
    year = "1995"
}

@article{Kim:2026mal,
    author = {Kim, Hyukjung and Kuzu, {\.I}layda and {\"O}zsoy, Kerem and Kahraman, Zeynep and Park, Wan-Il and Zoe, Heeseung},
    title = "{Phase transitions and gravitational wave production at the end of thermal inflation}",
    eprint = "2606.20392",
    archivePrefix = "arXiv",
    primaryClass = "hep-ph",
    doi = "10.1088/1475-7516/2026/09/120",
    journal = "JCAP",
    volume = "09",
    pages = "120",
    year = "2026"
}

@article{Easther:2008sx,
    author = "Easther, Richard and Giblin, Jr., John T. and Lim, Eugene A. and Park, Wan-Il and Stewart, Ewan D.",
    title = "{Thermal Inflation and the Gravitational Wave Background}",
    eprint = "0801.4197",
    archivePrefix = "arXiv",
    primaryClass = "astro-ph",
    reportNumber = "KAIST-TH-2008-01",
    doi = "10.1088/1475-7516/2008/05/013",
    journal = "JCAP",
    volume = "05",
    pages = "013",
    year = "2008"
}

@article{Linde:1980tt,
    author = "Linde, Andrei D.",
    title = "{Fate of the False Vacuum at Finite Temperature: Theory and Applications}",
    reportNumber = "LEBEDEV-80-92",
    doi = "10.1016/0370-2693(81)90281-1",
    journal = "Phys. Lett. B",
    volume = "100",
    pages = "37--40",
    year = "1981"
}

@article{Khlebnikov:1998sz,
    author = "Khlebnikov, S. and Kofman, L. and Linde, Andrei D. and Tkachev, I.",
    title = "{First order nonthermal phase transition after preheating}",
    eprint = "hep-ph/9804425",
    archivePrefix = "arXiv",
    reportNumber = "PURD-TH-98-06, UH-IFA-98-29, SU-ITP-98-29",
    doi = "10.1103/PhysRevLett.81.2012",
    journal = "Phys. Rev. Lett.",
    volume = "81",
    pages = "2012--2015",
    year = "1998"
}

@article{Jinno:2017ixd,
    author = "Jinno, Ryusuke and Lee, Sangjun and Seong, Hyeonseok and Takimoto, Masahiro",
    title = "{Gravitational waves from first-order phase transitions: Towards model separation by bubble nucleation rate}",
    eprint = "1708.01253",
    archivePrefix = "arXiv",
    primaryClass = "hep-ph",
    reportNumber = "CTPU-17-28, KEK-TH-1990",
    doi = "10.1088/1475-7516/2017/11/050",
    journal = "JCAP",
    volume = "11",
    pages = "050",
    year = "2017"
}

@article{Sui:2025epg,
    author = "Sui, Xiao-Bin and Liu, Jing and Cai, Rong-Gen",
    title = "{Curvature perturbations and gravitational waves induced by a first-order phase transition during reheating}",
    eprint = "2511.06421",
    archivePrefix = "arXiv",
    primaryClass = "gr-qc",
    doi = "10.1103/26jv-jczj",
    journal = "Phys. Rev. D",
    volume = "113",
    number = "8",
    pages = "083524",
    year = "2026"
}

@article{Lewicki:2024sfw,
    author = "Lewicki, Marek and Toczek, Piotr and Vaskonen, Ville",
    title = "{Black holes and gravitational waves from phase transitions in realistic models}",
    eprint = "2412.10366",
    archivePrefix = "arXiv",
    primaryClass = "astro-ph.CO",
    doi = "10.1016/j.dark.2025.102075",
    journal = "Phys. Dark Univ.",
    volume = "50",
    pages = "102075",
    year = "2025"
}

@article{Franciolini:2025ztf,
    author = "Franciolini, Gabriele and Gouttenoire, Yann and Jinno, Ryusuke",
    title = "{Curvature Perturbations from First-Order Phase Transitions: Implications to Black Holes and Gravitational Waves}",
    eprint = "2503.01962",
    archivePrefix = "arXiv",
    primaryClass = "hep-ph",
    reportNumber = "CERN-TH-2025-044, KOBE-COSMO-25-05",
    doi = "10.1103/tfcx-kzqx",
    journal = "Phys. Rev. Lett.",
    volume = "136",
    number = "17",
    pages = "171404",
    year = "2026"
}

@article{Wang:2026zvz,
    author = "Wang, Xiao and Bal{\'a}zs, Csaba and Ding, Ran and Tian, Chi",
    title = "{How large are curvature perturbations from slow first-order phase transitions? A gauge-invariant analysis}",
    eprint = "2601.14412",
    archivePrefix = "arXiv",
    primaryClass = "hep-ph",
    doi = "10.1103/y9ks-n9sv",
    journal = "Phys. Rev. D",
    volume = "113",
    number = "12",
    pages = "123026",
    year = "2026"
}

@article{Bae:2025vkn,
    author = "Bae, Jeong-Myeong and Mohammad, Hammam Raihan and Stewart, Ewan D. and Zoe, Heeseung",
    title = "{The curvature perturbation generated by thermal fluctuations during thermal inflation}",
    eprint = "2501.06580",
    archivePrefix = "arXiv",
    primaryClass = "astro-ph.CO",
    doi = "10.1088/1475-7516/2025/04/038",
    journal = "JCAP",
    volume = "04",
    pages = "038",
    year = "2025"
}

\end{document}